\documentclass[5p]{elsarticle}

\usepackage{lineno,hyperref}
\modulolinenumbers[5]

\usepackage{multirow}
\usepackage{tabularx}
\usepackage{booktabs}
\usepackage{framed}
\usepackage{hyperref}
\usepackage{subcaption}
\usepackage{graphicx}
\usepackage[table]{xcolor}

\newcommand{\numPapers}{139}

\definecolor{lightgray}{gray}{0.9}

\journal{Journal of Systems and Software}

\begin{document}
\begin{frontmatter}

\title{A Systematic Literature Review and Taxonomy of Modern Code Review}

\author{Nicole Davila\corref{cor1}}
\ead{ncdavila@inf.ufrgs.br}
\cortext[cor1]{Corresponding author.}

\author{Ingrid Nunes}
\ead{ingridnunes@inf.ufrgs.br}

\address{Universidade Federal do Rio Grande do Sul (UFRGS), Instituto de Inform\'atica, Porto Alegre, Brazil}

\begin{abstract}
\textbf{Context:} Modern Code Review (MCR) is a widely known practice of software quality assurance. However, the existing body of knowledge of MCR is currently not understood as a whole. \textbf{Objective:} Our goal is to identify the state of the art on MCR, providing a structured overview and an in-depth analysis of the research done in this field. \textbf{Method:} We performed a systematic literature review, selecting publications from four digital libraries. \textbf{Results:} A total of \numPapers{} papers were selected and analyzed in three main categories. \textsc{Foundational studies} are those that analyze existing or collected data from the adoption of MCR. \textsc{Proposals} consist of techniques and tools to support MCR, while \textsc{evaluations} are studies to assess an approach or compare a set of them. \textbf{Conclusion:} The most represented category is \textsc{foundational studies}, mainly aiming to understand the motivations for adopting MCR, its challenges and benefits, and which influence factors lead to which MCR outcomes. The most common types of \textsc{proposals} are code reviewer recommender and support to code checking. \textsc{Evaluations} of MCR-supporting approaches have been done mostly offline, without involving human subjects. Five main research gaps have been identified, which point out directions for future work in the area.
\end{abstract}


\end{frontmatter}

\section{Introduction}\label{sec:introduction}

Code review is a widely known practice of software quality assurance. It consists of developers, other than the author, manually checking code changes before they are integrated into the main code repository. The goal is to look for defects or improvement opportunities without the software execution and before the product delivery, thus reducing the costs of fixing them later~\citep{ID414:BaumEtAl2016}. Due to the   adoption of agile methods and distributed software development, code reviews are currently done in a less formal way than in the past, reducing the inefficiencies of its early form, namely software inspections~\citep{Fagan1976}. The lightweight variant of code review has been referred to as \emph{modern code review} (MCR)~\citep{ID105:BacchelliBird2013}, which is a flexible, tool-based, and asynchronous process. The focus is on small code changes, and reviews happen early, quickly, and frequently, which helps detect defects and problem-solving, among other benefits~\citep{ID116:RigbyBird2013}. Moreover, because MCR is mediated by tools, it is possible to leverage MCR databases to learn from them and build complementary tools~\citep{ID464:ThongtanunamEtAl2015, ID600:OuniEtAl2016}.

The increasing popularity of MCR in open source and industrial projects motivated studies on the practice, resulting in a large and also increasing body of knowledge in the literature. Due to this, it is a challenge to identify the main contributions, difficulties, and research opportunities related to MCR. Therefore, it is important to gather and review the existing contributions to be able to understand the state of the art and guide future efforts.

Our goal in this study is thus to obtain a comprehensive overview of the literature on MCR, which is helpful for researchers and practitioners to understand what has already been explored, open questions, and lessons learned about the practice. To achieve this goal, we present the results of a broad and in-depth systematic literature review of research work on MCR. Following a systematic method to search, select, and analyze relevant studies, \numPapers{} (out of 825 retrieved) papers have been investigated. We focus on answering three research questions, each related to a different type of study. Each type corresponds to one of the three main typical types of research contributions in this context. \textsc{Foundational studies} are those that analyze existing or collected data related to MCR, exploring various aspects of the practice and deriving knowledge to improve it. From these studies, we discuss their key findings, which consist mainly of evidence about the practice in real settings, such as common characteristics, influencing factors, process variants, and impact of the practice. \textsc{Proposals} consist of proposed approaches, such as techniques and tools, to support MCR. Learning the proposals that are currently available allows researchers to understand and address limitations of existing support to MCR and practitioners to know what is available for them to improve the MCR practice in particular projects. Lastly, \textsc{evaluations} are studies to assess a proposed approach or compare a set of them. We compile how existing approaches compare to each other, thus providing evidence of their effectiveness. Moreover, this analysis also provides guidance for researchers to understand how to evaluate their work on MCR. Furthermore, based on our collected and analyzed data, we derive a new taxonomy of MCR research and identify issues that remain unaddressed. This paves the way for future work on MCR, which is a practice that has been largely used to improve the quality of software systems.

There are previous secondary studies on MCR. Three systematic mapping studies were done on this topic. Two \citep{MappingMCR:FronzaEtAl2020,MappingMCR:BadampudiEtAl2019} aimed at identifying themes that are typically target of research, while that made by \citet{MappingRefactoringAware:CoelhoEtAl2019} focused on refactoring-aware code review. In addition, there are systematic literature reviews on MCR with a narrower scope than ours. They cover the MCR benefits~\citep{MappingMCR:NazirEtAl2020}, the individual elements that impact on knowledge sharing~\citep{SLRKnowledgeSharing:FatimaEtAl2019,SLRFeedbackKnowledge:FatimaEtAl2019}, and the situational variables influencing sustainability in MCR~\citep{SLRSituationalFactors:NazirEtAl2019,SLRSituationalFactors2:NazirEtAl2019}. In addition, \citet{InfSLRrecommenders:DouganEtAl2019} conducted a narrative literature review to investigate the validity of ground truth data in reviewer recommender studies. Although these secondary studies are relevant contributions to the field, our review takes a step further by making a broader and in-depth analysis of the literature and capturing a structured view of their key findings.

We next provide background on MCR. The methodology adopted to perform our systematic literature review is presented in Section~\ref{sec:Method}, followed by the analysis of its results that are described in Sections~\ref{sec:Foundational}--\ref{sec:Evaluation}. Insights derived from our study and its limitations are discussed in Section~\ref{sec:Discussion}. In Section~\ref{sec:Conclusion}, we conclude.


\section{Background on Modern Code Review}
\label{sec:Background}

The term \emph{modern code review} (MCR) was first used by \citet{MCR:Cohen2010} and became popular by the study of \citet{ID105:BacchelliBird2013}. There are no strict guidelines for adopting MCR, causing it to be a flexible practice. The overall idea is that developers other than the author---i.e.\ the reviewers---assess code changes to identify both defects and quality problems and decide whether the changes will be discarded or integrated into the main project repository. MCR is mainly characterized by being asynchronous and supported by tools, occurring regularly integrated with the routine of developers~\citep{ID105:BacchelliBird2013,ID127:SadowskiEtAl2018}. 

\begin{figure}[t]
 \centering
 \includegraphics[width=0.9\linewidth]{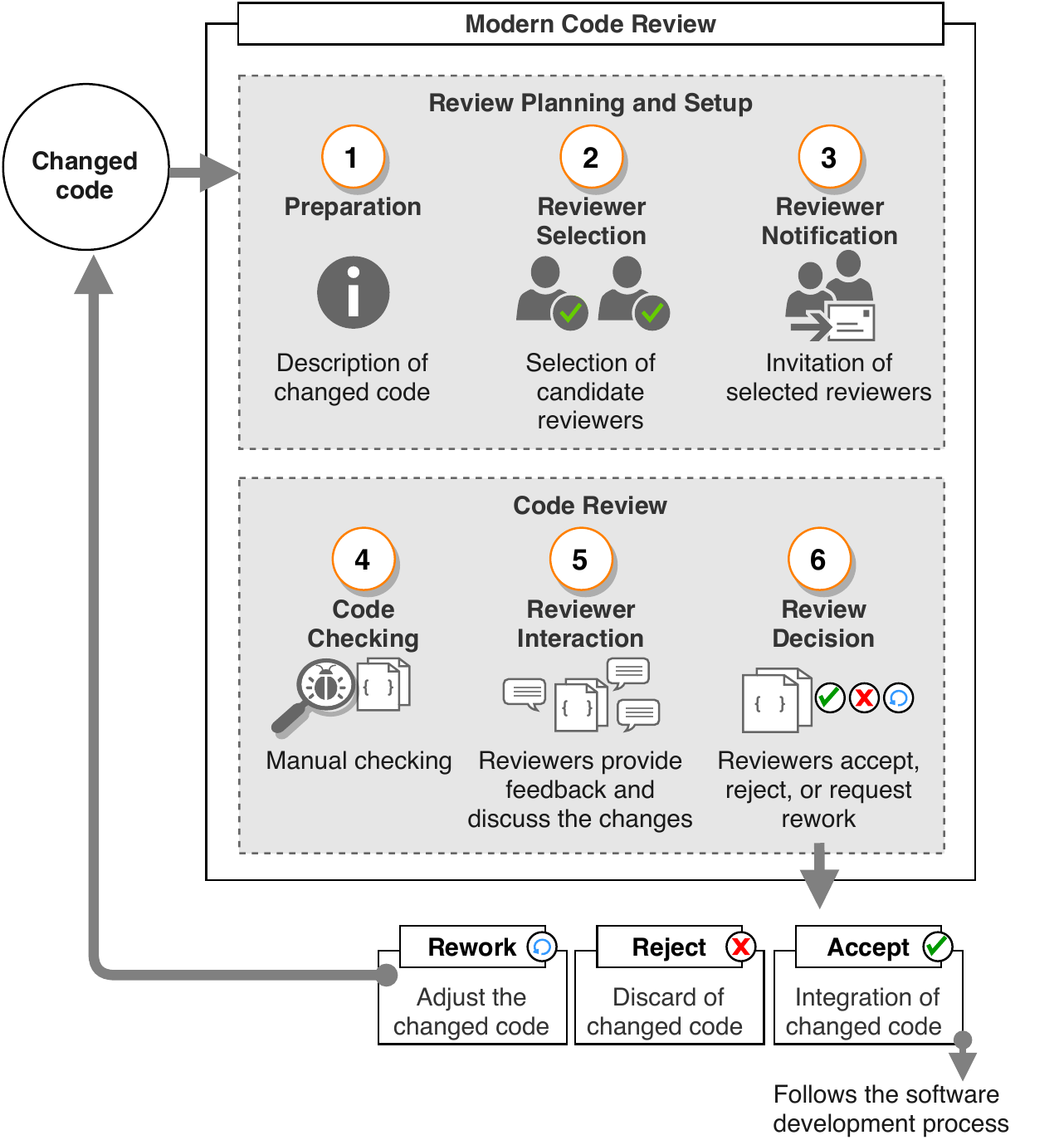}
 \caption{An overview of the tasks of modern code review.}
 \label{fig:overviewMCR}
\end{figure}

To understand how MCR generally works, we summarize in Figure~\ref{fig:overviewMCR} a typical MCR process, capturing tasks performed to review a code change using well-known code review tools, such as Gerrit and the pull-request system of GitHub. Each organization or project can have a customized MCR process with particular tools, policies, and culture. Figure~\ref{fig:overviewMCR} splits the MCR tasks into two phases, \emph{Review Planning and Setup} and \emph{Code Review}, where the former consists of tasks to enable the review to take place, while the latter is the phase when the code is assessed and decisions based on the review are made. These tasks are performed by at least two roles, namely author and reviewer. The author is a developer who changes the source code and creates a review request for it, while the reviewer (typically also developers) is responsible for reviewing the code and giving feedback. In particular settings, there might be other roles, such as maintainer~\citep{ID820:SantosNunes2018} (a developer responsible for a system's module and is required to approve code changes in it) and commenter~\citep{ID079:JiangEtAl2017} (who can provide comments but not make decisions). 

The first phase, \emph{Review Planning and Setup}, consists of three main tasks. In the first (\emph{Preparation}), the author, who has a unit of work to be reviewed, prepares a review package composed of the changed code accompanied by a description providing additional details of the change. In the \emph{Reviewer Selection} task, suitable reviewers that likely can or should inspect the review package are selected. Finally, in the \emph{Reviewer Notification} task, reviewers receive a notification inviting them to perform the review. 

In the second phase, \emph{Code Review}, the process of analyzing the code for defects and improvement opportunities takes place. Each reviewer individually performs \emph{Code Checking}, assessing changes and comparing it with previous versions. They also interact among themselves and with the author (\emph{Reviewer Interaction}), writing feedback comments, annotating snippets of code, or promoting a discussion to clarify issues. Based on this interaction, reviewers make a decision on the review request (\emph{Review Decision}), which can be \emph{accept}, \emph{reject}, and \emph{rework}. The last results in a new review cycle to take place after the code is updated according to the reviewer feedback.

How each MCR task occurs in a concrete way depends on the selected supporting tools and customization made in particular projects. Tools can be used to, e.g., manage review requests, select reviewers, visualize and analyze code changes, register annotations, and manage the discussions. Additional tools can be used to automatically assess code changes, check for standards, collect metrics, indicate potential bugs, and so on. Feedback can be given in the form of comments or votes, and a vote can be done in different scales.

The flexibility of MCR allows it to be employed by teams and organizations in various settings using different technologies. It has been used in GitHub to review pull-requests and in industry, assisted by proprietary software with code review features or well-known code review tools, such as Gerrit. The MCR flexibility and its tailored adoption in different contexts motivated research to identify and understand this practice as well as improve it. This led to many studies, which are surveyed in this work.


\section{Systematic Literature Review} \label{sec:Method}

\begin{figure}[t]
 \centering
 \includegraphics[width=\linewidth]{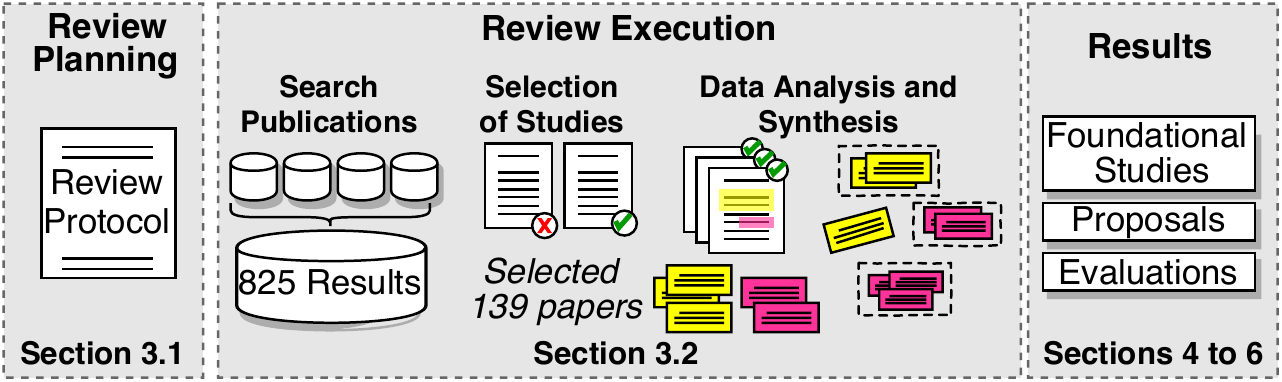}
 \caption{Main steps of our systematic literature review.}
 \label{fig:methodology}
\end{figure}

To perform our systematic literature review (SLR), we followed the guidelines proposed by~\citet{SLR:KitchenhamCharters2007}. We executed three main steps---Review Planning, Review Execution, and Data Analysis and Synthesis---which are illustrated in Figure~\ref{fig:methodology}. This section details the protocol followed to conduct our SLR and the results of the literature search and selection.

\subsection{Systematic Review Planning}\label{sec:protocol}

\subsubsection{Research Goals} 

Our goal is to obtain a comprehensive overview of the literature on MCR. Over the past years, this topic received significant research interest, which led to many studies. Our goal is to gather, structure, and compare studies' findings. Such an in-depth analysis of research on MCR is helpful for researchers to understand what has already been explored and open questions. It is also useful for practitioners, who can learn how to achieve better outcomes with MCR by identifying how to adjust the MCR practice in particular settings and becoming aware of existing approaches that can be adopted to support the practice. 

\subsubsection{Research Questions}
Based on our goal, our main research question is: \emph{what is the state of the art of research on modern code review?} Considering this broad research question, we focus on different types of research work, covering three typical research directions: \textsc{foundational studies}, which are those that analyze existing or collected data associated with MCR to gather knowledge about the practice that provide practitioners with knowledge to improve the adoption of the practice or researchers with findings to build further research on them; \textsc{proposals}, which consist of novel techniques and tools to support MCR, and \textsc{evaluations}, which are studies to assess a proposed approach or compare a set of them. This leads us to the following specific research questions (RQ).

\begin{itemize}
\item \textbf{RQ-1}: What foundational body of knowledge has been built based on studies of MCR?
\item \textbf{RQ-2}: What approaches have been developed to support MCR?
\item \textbf{RQ-3}: How have MCR approaches been evaluated and what were the reached conclusions? 
\end{itemize}

\subsubsection{Search Strategy}

To find the studies that are relevant to our work, we selected four databases, shown in Table~\ref{tab:searchResults}, which store papers published in many key software engineering conferences and journals. Our search does not include other databases, e.g.\ Google Scholar and arXiv, because they provide pointers to papers already stored in our searched databases or include non-peer-reviewed papers. We focus on papers that are peer reviewed as a means to obtain evidence of the quality of their research.

\begin{table}
\caption{Search results by source.}
\label{tab:searchResults}
\centering
\begin{tabularx}{\linewidth}{p{2.5cm} X r}
\toprule
\textbf{Databases}                & \textbf{URL}& \textbf{\#Studies} \\ 
\midrule
ACM DL                             & http://portal.acm.org        & 291 \\
IEEE Xplore                        & http://ieeexplore.ieee.org   & 394 \\
ScienceDirect                     & http://www.sciencedirect.com & 122 \\
SpringerLink                      & http://link.springer.com     & 76  \\
\midrule
Duplicates                        && 58                \\
\midrule
\multicolumn{2}{l}{\textbf{Total (including duplicates)}}& \textbf{883} \\
\multicolumn{2}{l}{\textbf{Total (excluding duplicates)}}& \textbf{825} \\ 
\bottomrule
\end{tabularx}
\end{table}

To retrieve the publications from the selected databases, we used as keyword the term \emph{code review} and the following synonyms: \emph{code inspection}, \emph{software inspection}, and \emph{formal inspection}. Though our goal is to identify work on MCR, some authors use the term inspection but refer to a lightweight process. We, therefore, identified whether papers focus on MCR not by the adopted term but by their description of the practice. The term \emph{code review} is a substring of other synonyms, e.g.\ modern code review, changed-based code review, contemporary peer code review, and peer code review. Thus, there is no need to include them. The only term not covered, which was used by a few authors, is \emph{peer review} because it leads to a huge amount of papers related neither to MCR nor to Computer Science and would make the SLR infeasible to be done in a timely manner. Our resulting search string is as follows.
\begin{framed}
\noindent\textbf{Search String}: ``code review'' OR ``code inspection'' OR ``software inspection'' OR ``formal inspection''
\end{framed}

\subsubsection{Selection Criteria}

The papers retrieved by querying search databases are filtered using selection criteria,  summarized in Table~\ref{tab:criteria}. To be selected, a primary study must satisfy at least one inclusion criterion and no exclusion criterion.

\begin{table}
\caption{Inclusion criteria (IC) and exclusion criteria (EC).}
\label{tab:criteria}
\centering
\begin{tabularx}{\linewidth}{l X}
\toprule
\multicolumn{2}{l}{\textbf{Criteria}} \\ 
\midrule
IC-1                  & The paper presents a study of foundational aspects of MCR. \\
IC-2                  & The paper proposes an approach to support MCR.  \\
IC-3                  & The paper presents an evaluation of an MCR approach. \\
\midrule
EC-1                  & The paper does not focus on MCR as a reviewing practice made by peers within the software development. \\ 
EC-2                  & The paper is not written in English. \\
EC-3                  & The content of the paper was also published in another more complete paper, which is already included. \\
EC-4                  & We have no access to the full paper. \\
EC-5                  & The content is not a scientific paper of at least four pages. \\
\bottomrule
\end{tabularx}
\end{table}
 
Our EC-1 excludes studies that use MCR as a motivation or as an example of context where an approach can be applied. Their focus is on techniques that can be incorporated to MCR. For example, this is the case of static analysis approaches that can be used by reviewers or as part of an automated reviewer but can also be applied to any context to automatically analyze the source code. This EC also excludes studies that focus on code review being used for purposes other than software development, such as teaching software engineering.

\subsection{Review Execution} \label{sec:execution}

The execution of the review protocol detailed in the previous section led to the results described as follows.

\subsubsection{Search Execution}

Our search string was customized according to the specific syntax of each search engine. We searched within the abstracts of the publications. Due to limitations in the Springer Link database, which does not allow searches in abstracts, we considered paper keywords instead. We searched for papers published before or in 2019. The number of retrieved papers is shown in Table \ref{tab:searchResults}.

\subsubsection{Selection of Primary Studies}

To select primary studies, we first selected papers that matched any inclusion criterion, based on their title and abstract. Then, in a second step, we evaluated all inclusion and exclusion criteria based on the full text of the paper. Table~\ref{tab:selectionResults} depicts the results of the execution of these steps. After a precise specification of each IC and EC and the joint analysis of papers for verifying a common understanding of their meaning, each primary study was analyzed by a single researcher. If the researcher could not  evaluate the satisfaction of any of the criteria, the opinion of the second researcher was requested to minimize the potential researcher bias. If there was no agreement, both researchers discussed until they converged to a decision.

\begin{table}\caption{Selection of studies based on inclusion and exclusion criteria. In the full analysis, 33 papers matched two IC and two papers matched three IC.}
\label{tab:selectionResults}
\centering
\begin{tabular}{l rrrr}
\toprule
\textbf{Step/Criterion}   & \textbf{IC-1} & \textbf{IC-2} & \textbf{IC-3} & \textbf{Total} \\ \hline
Abstract analysis         & 228                   & 146               & 45                  & 378               \\
Full analysis             & 98                    & 60                & 37                  & 158               \\ \midrule
EC-1                      &                       &                   &                     & 0                 \\
EC-2                      &                       &                   &                     & 0                 \\
EC-3                      & 3                     & 5                 &                    & 8                 \\
EC-4                      &                       &                   &                    & 0                 \\
EC-5                      & 9                     & 2                  &                    & 11                 \\ \hline
Selected studies & \textbf{86}           & \textbf{53}       & \textbf{37}      & \textbf{139}      \\ 
\bottomrule
\end{tabular} 
\end{table}

Papers that do not focus on MCR do not match any IC. For example, there are studies excluded because they focus on software inspection as a synchronous process with an inspection meeting. Moreover, there are papers satisfying multiple IC, e.g.~\citep{ID207:RahmanEtAl2017,ID210:ThongtanunamEtAl2014}, which describe a combination of foundational studies (IC-1), proposed approach (IC-2), and evaluation (IC-3). In addition, there are papers discarded due to the satisfaction of ECs. Due to EC-1, we excluded studies of pedagogical code review~\citep{Pedagogical:HundhausenEtAl2011,Pedagogical:PetersenZingaro2018} and static analysis tools~\citep{StaticAnalyzers:BurhandennyEtAl2016,StaticAnalyzers:SinghEtAl2017}. As said, automated source code analysis can be used in other contexts and, if we had not excluded these studies, any approach that performs static analysis should have been included.

\begin{figure}[!ht]
 \centering
 \includegraphics[width=0.9\linewidth]{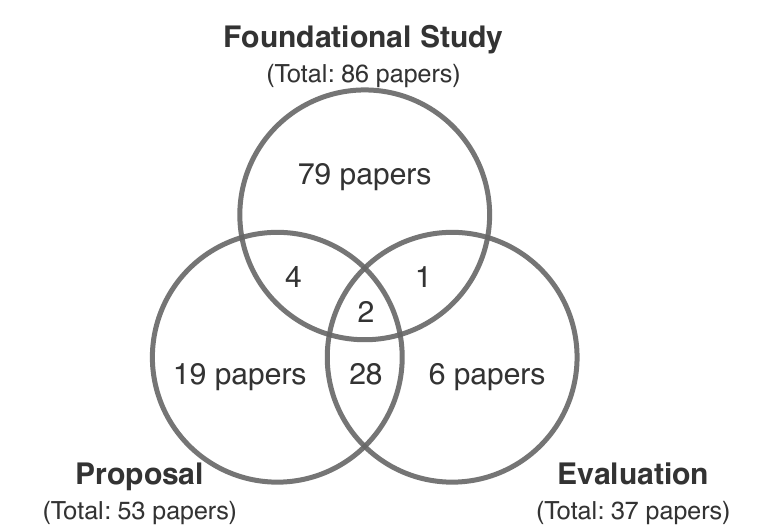}
 \caption{Distribution of primary studies in three main categories and their intersections.}
 \label{fig:diagramPrimaryStudies}
\end{figure}

Our final set of selected primary studies has a total of \numPapers{} publications. Papers are categorized according to the IC that they satisfy: (i) \textsc{foundational studies} investigate aspects of MCR or an experience report of the use of MCR (IC-1); (ii) \textsc{proposals} describe novel approach to support MCR, such as a tool, technique or method (IC-2); and  (iii) \textsc{evaluations} of MCR approaches (IC-3). Throughout the paper, the highlighted terms are used to refer to these study types. Although the majority of papers present a single type of contribution, 35 include two or three types of studies and are therefore in multiple categories, as shown in the Venn diagram in Figure~\ref{fig:diagramPrimaryStudies}. Table~\ref{tab:finalSetStudies} shows the papers selected for analysis grouped by category. 

\begin{table*}[t]
\caption{Selected primary studies.}
\label{tab:finalSetStudies}
\footnotesize
\centering
\begin{tabularx}{\linewidth}{X}
\toprule
\textsc{Foundational Studies} \\ 
\midrule
\citet{ID077:Muller2005,ID103:McIntoshEtAl2014,ID104:ShimagakiEtAl2016,ID105:BacchelliBird2013,ID107:AlbayrakDavenport2010,ID110:BellerEtAl2014,ID116:RigbyBird2013,ID125:GermanEtAl2018,ID126:RahmanRoy2017,ID127:SadowskiEtAl2018,ID131:BosuEtAl2015,ID134:ThompsonWagner2017,ID140:YangEtAl2017,ID144:KitagawaEtAl2016,ID147:Izquierdo-CortazarEtAl2017,ID149:BosuCarver2012,ID154:SpadiniEtAl2018,ID156:BaumEtAl2016,ID158:BirdEtAl2015,ID161:KononenkoEtAl2016,ID171:BosuCarver2014,ID172:ThongtanunamEtAl2015,ID191:KovalenkoBacchelli2018,ID202:EfstathiouSpinellis2018,ID206:BosuEtAl2014,ID207:RahmanEtAl2017,ID209:BegelVrzakova2018,ID211:DunsmoreEtAl2000,ID255:HiraoEtAl2015,ID267:ThongtanunamEtAl2016,ID277:FloydEtAl2017,ID291:UwanoEtAl2006,ID318:LeeCarver2017,ID331:MeneelyEtAl2014,ID365:BosuEtAl2017,ID398:SutherlandVenolia2009,ID408:ChandrikaEtAl2017,ID410:AsundiJayant2007,ID427:MacLeodEtAl2018,ID449:UedaEtAl2017,ID450:ArmstrongEtAk2017,ID464:ThongtanunamEtAl2015,ID465:EbertEtAl2017,ID480:KononenkoEtAl2015,ID484:BavotaRusso2015,ID487:RunesonAndrews2003,ID509:BosuCarver2013,ID511:LiangMizuno2011,ID532:BiaseEtAl2016,ID538:PanichellaEtAl2015,ID560:SwamiduraiEtAl2014,ID590:MoralesEtAl2015,ID594:BaysalEtAl2012,ID595:RigbyEtAl2012,ID599:BernhartGrechenig2013,ID613:DuraesEtAl2016,ID624:MurakamiEtAl2017,ID642:LiEtAl2017,ID644:ThongtanunamEtAl2017,ID645:BaumEtAl2017,ID651:McIntoshEtAl2016,ID653:BaysalEtAl2016,ID656:HiraoEtAl2016,ID704:AlamiEtAl2019,ID705:PascarellaEtAl2018,ID709:SpadiniEtAl2019,ID710:ZanatyEtAl2018,ID712:HiraoEtAl2019,ID713:RamEtAl2018,ID746:EbertEtAl2018,ID750:AnEtAl2018,ID758:UedaEtAl2018,ID759:NorikaneEtAl2018,ID767:UedaEtAl2019,ID768:ZampettiEtAl2019,ID770:EbertEtal2019,ID771:JiangEtAl2019,ID778:PaixaoMaia2019,ID783:PaulEtAl2019,ID785:PaixaoEtAl2019,ID807:ElAsriEtAl2019,ID812:WangEtAl2019,ID815:BaumEtAl2019,ID816:RuangwanEtAl2019,ID820:SantosNunes2018} \\
\midrule
\textsc{Proposals} \\ 
\midrule
\citet{ID051:YuEtAl2016,ID079:JiangEtAl2017,ID131:BosuEtAl2015,ID132:MullerEtAl2012,ID138:Balachandran2013,ID146:BarnettEtAl2015,ID155:RahmanEtAl2016,ID175:ZhangEtAl2015,ID195:SripadaEtAl2016,ID207:RahmanEtAl2017,ID210:ThongtanunamEtAl2014,ID224:HaoEtAl2013,ID234:TaoKim2015,ID235:PriestPlimmer2006,ID258:SoltanifarEtAl2016,ID276:DuleyEtAl2010,ID291:UwanoEtAl2006,ID378:KalyanEtAl2016,ID400:GeEtAl2017,ID401:ZanjaniEtAl2016,ID405:TymchukEtAl2015,ID409:AhmedEtAl2017,ID413:BaumElAl2017,ID414:BaumEtAl2016,ID425:NagoyaEtAl2005,ID434:LanubileMallardo2002,ID439:HarelKantorowitz2005,ID464:ThongtanunamEtAl2015,ID465:EbertEtAl2017,ID486:PangsakulyanontEtAl2014,ID493:XiaEtAl2017,ID572:ZhangEtAl2011,ID598:MishraSureka2014,ID600:OuniEtAl2016,ID601:MenariniEtAl2017,ID608:PerryEtAl2002,ID614:WangEtAl2017,ID616:XiaEtAl2015,ID637:Aman2013,ID638:FejzerEtAl2018,ID639:FanEtAl2018,ID641:FreireEtAl2018,ID642:LiEtAl2017,ID655:BaumSchneider2016,ID706:AsthanaEtAl2019,ID711:HuangEtAl2018,ID716:WangEtAl2019,ID736:HuangEtAl2018,ID753:WenEtAl2018,ID761:HanamEtAl2019,ID772:LiaoEtAl2019,ID803:JiangEtAl2019,ID817:GuoEtAl2019}\\
\midrule
\textsc{Evaluations} \\
\midrule
\citet{ID051:YuEtAl2016,ID102:KhandelwalEtAl2017,ID132:MullerEtAl2012,ID138:Balachandran2013,ID140:YangEtAl2017,ID146:BarnettEtAl2015,ID155:RahmanEtAl2016,ID175:ZhangEtAl2015,ID207:RahmanEtAl2017,ID210:ThongtanunamEtAl2014,ID234:TaoKim2015,ID257:BaumEtAl2016,ID276:DuleyEtAl2010,ID281:PengEtAl2018,ID298:HannebauerEtAl2016,ID400:GeEtAl2017,ID401:ZanjaniEtAl2016,ID464:ThongtanunamEtAl2015,ID493:XiaEtAl2017,ID600:OuniEtAl2016,ID601:MenariniEtAl2017,ID616:XiaEtAl2015,ID637:Aman2013,ID638:FejzerEtAl2018,ID639:FanEtAl2018,ID641:FreireEtAl2018,ID661:MizunoLiang2015,ID698:RunesonWohlin1998,ID706:AsthanaEtAl2019,ID711:HuangEtAl2018,ID716:WangEtAl2019,ID736:HuangEtAl2018,ID753:WenEtAl2018,ID761:HanamEtAl2019,ID772:LiaoEtAl2019,ID803:JiangEtAl2019,ID817:GuoEtAl2019}\\
\bottomrule
\end{tabularx}
\end{table*}

\subsubsection{Data Extraction and Information Labeling}

\begin{figure}[!ht]
 \centering
 \includegraphics[width=0.97\linewidth]{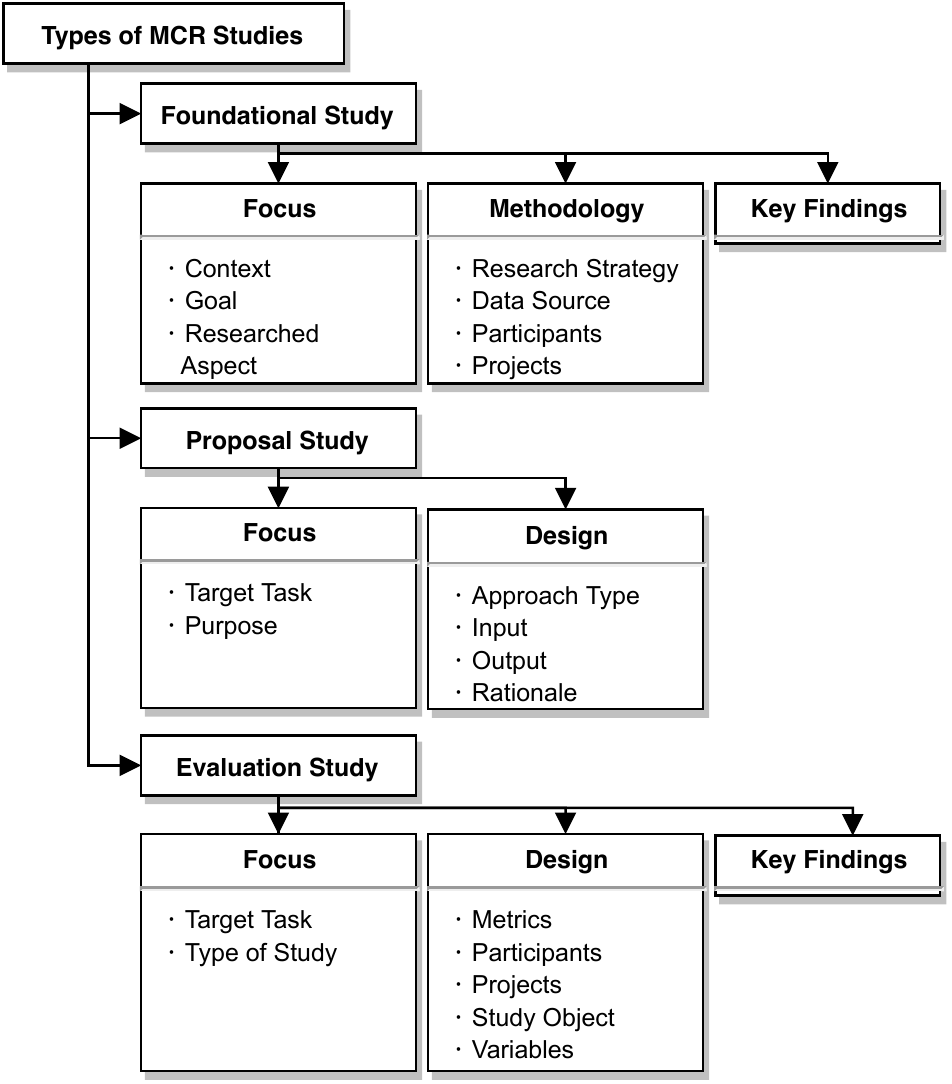}
 \caption{Facets analyzed in each primary study.}
 \label{fig:analysisFacets}
\end{figure}

To answer each research question, we extracted information from each study according to the facets indicated in Figure~\ref{fig:analysisFacets}. If a paper includes multiple types of study, each study of the paper was analyzed separately.

To label the information collected from the studies associated with each facet, we used coding. \emph{Coding} is used to derive a framework of thematic ideas of qualitative data (text or images) by indexing data~\citep{Gibbs:chapter2007:Coding}. All studies were initially broadly analyzed to identify an initial set of codes. Then, studies were analyzed in-depth, with the codes being refined in the process. Finally, some codes were merged (when they conveyed the same underlying idea), resulting in our final set of codes. A classification was jointly elaborated by the two authors of this paper, with definitions as precise as possible of each classification to be made. To maintain consistency in the classification, the studies of each main category, i.e., \textsc{foundational studies}, \textsc{proposals}, and \textsc{evaluations}, were carefully read, analyzed, and coded by a single researcher. In cases in which there was no certainty while classifying the paper content based on the definitions, the two researchers analyzed the study and discussed until they converged on the codes that should be used. From the 139 primary studies, there were 27 (approximately 20\%) cases in which the first author faced uncertainty in the classification. Considering the \textsc{foundational studies}, the researchers had five interactions until they reached a final version of the codes. There were four refinement interactions for \textsc{proposals}, and for \textsc{evaluations}, there was the need for only two interactions. For example, considering de refinement of aspects related to \textsc{foundational studies}, we had in the set of codes the labels \emph{Analysis of the Impact on Internal Outcomes} and \emph{Analysis of Activities Outcomes}, which after the refinements were merged into \emph{Analysis of Internal Outcomes} because both included studies targeting internal outcomes.

The coding process was performed at different levels in each main category, as summarized in Figure~\ref{fig:analysisFacets}. We used coding to categorize information of the study focus, methodology, and key findings of \textsc{foundational studies}. In the \textsc{proposals} category, we coded the approach focus and its design. Then, studies in the \textsc{evaluation} group were labeled considering their focus, design, and key findings. 

\subsubsection{Findings of the Systematic Literature Review}

The next three sections present our findings by groups of studies (\textsc{foundational studies}, \textsc{proposals}, and \textsc{evaluations}), which are associated with our three research questions, respectively.\footnote{The complete analysis of each individual primary study can be found online at \url{http://inf.ufrgs.br/prosoft/resources/2020/jss-code-review-slr}.} We first provide an overview of the category and then we discuss the subcategories in detail. Throughout the discussion, we highlighted the key findings of each category using frames. At the end of each section we answer the respective research question.


\section{\textsc{Foundational Studies}} \label{sec:Foundational}

\begin{table*}[t]
\caption{Taxonomy of topics researched in \textsc{foundational studies}.}
\label{tab:Results-Foundational-Codes}
\centering
\small
\rowcolors{1}{}{lightgray}
\begin{tabularx}{\linewidth}{p{2.8cm} X p{6.5cm} r}
\toprule
\textbf{Category} & \textbf{Main Goal} & \textbf{Typical Research Method} & \textbf{\#Studies}\\ 
\midrule
Practitioner{\par}Perception of the Practice & Learn from a developers' perspective the state of the practice of MCR & Collection and analysis of subjective or qualitative data with the opinion and thoughts of the practitioners & 18 \\
Analysis of the{\par}Innerworkings of the MCR Process & Analyze the reviewing process in real settings, sharing of insights and lessons learned of the practice adoption & Contributions based on quantitative analysis, experience or the analysis of a case study & 6 \\
Cross-patch{\par}Analysis & Analyze the relationship among reviewed patches and identify recurring patterns & Collection and analysis of data from code review traces and software development history & 3 \\
Analysis of{\par}Internal Outcomes & Understand key properties and metrics associated with the MCR process and what influences them & Collection and analysis of data of code in repositories or collected in experiments & 39\\
Analysis of{\par}External Outcomes & Understand how MCR influences outcomes  external to the review process (code and product quality) & Collection and analysis of data from repositories of code review, version control systems and issue trackers & 13\\
Human Aspects of Reviewers  & Understand human behavior and individual aspects of the reviewers & Simulations or experiments to collect human body data (e.g., brain and eye activity) & 9\\
External Aspects of MCR & Analyze of the relationship between MCR and other software development practices & Experiments or data analysis from software development history & 8\\
\bottomrule
\end{tabularx}
\end{table*}

We classified the 86 \textsc{foundational studies} present in the selected papers into seven categories, as presented in Table~\ref{tab:Results-Foundational-Codes}. Most of the studies focus solely on MCR, while eight investigate the relationship of the practice with other approaches of software development. Studies classified as \emph{practitioner perception of the practice} and \emph{analysis of innerworkings of MCR process} analyze the state of the practice, exploring multiple aspects and sharing lessons learning from studies in real settings. Studies focusing on \emph{cross-patch analysis} search for links and patterns related to multiple reviewed patches. The other two categories---\emph{analysis of internal outcomes} and \emph{analysis of external outcomes}---investigate specific characteristics of the MCR process by means of data analysis. A smaller set of studies explores simulations or human body data (e.g.,\ brain and eye activity) to understand \emph{human aspects of reviewers}.  Finally, there are papers that focus on the relationship of MCR with other practices, which are classified as \emph{external aspects of MCR}. Given this overview of the categories of \textsc{foundational studies}, we next analyze each of them.

\subsection{Practitioner Perception of the Practice}

We identified in the literature 18 \textsc{foundational studies} that explore multiple perspectives of the MCR practice collecting the perceptions of the practitioners. In most studies, the opinion and thoughts of practitioners were collected by means of questionnaires~\citep{ID125:GermanEtAl2018,ID161:KononenkoEtAl2016,ID365:BosuEtAl2017,ID398:SutherlandVenolia2009, ID509:BosuCarver2013,ID599:BernhartGrechenig2013,ID645:BaumEtAl2017,ID770:EbertEtal2019,ID820:SantosNunes2018}, semi-structured interviews~\citep{ID131:BosuEtAl2015,ID154:SpadiniEtAl2018,ID156:BaumEtAl2016,ID704:AlamiEtAl2019,ID713:RamEtAl2018}, or both~\citep{ID127:SadowskiEtAl2018,ID709:SpadiniEtAl2019}. Moreover, some studies complemented these data using observation~\citep{ID105:BacchelliBird2013,ID427:MacLeodEtAl2018} or  meeting with practitioners~\citep{ID704:AlamiEtAl2019}, or deployed a specific tool to ask professionals for feedback~\citep{ID713:RamEtAl2018}. Half of these \textsc{foundational studies} involve participants from both industry and open-source projects. The remaining studies on perceptions of the practice involved only participants from open-source software (OSS) projects~\citep{ID125:GermanEtAl2018,ID161:KononenkoEtAl2016,ID509:BosuCarver2013,ID704:AlamiEtAl2019} or only participants from the industry~\citep{ID105:BacchelliBird2013,ID127:SadowskiEtAl2018,ID131:BosuEtAl2015,ID156:BaumEtAl2016,ID645:BaumEtAl2017,ID820:SantosNunes2018}.

The studies that capture the practitioner perception of the practice focus on four different MCR aspects, namely (i) \emph{reviewing practices}; (ii) \emph{social aspects}; (iii) \emph{expected benefits}; and (iv) \emph{challenges and difficulties}. The majority of the studies target only one aspect, while six of them~\citep{ID105:BacchelliBird2013, ID127:SadowskiEtAl2018, ID156:BaumEtAl2016, ID161:KononenkoEtAl2016, ID365:BosuEtAl2017, ID427:MacLeodEtAl2018} study multiple aspects. We next detail the findings provided by these studies.

\begin{framed}
\noindent\textbf{Finding 1}: The perception of the MCR is collected mainly through questionnaires (13 of 18 studies, approximately 72\%) and provides evidence of practitioners' opinions from industry (50\% of the studies), open-source projects (22\%), or both (28\%).
\end{framed}

\paragraph{\textbf{Reviewing Practices}}

More than a half of the studies that collected the opinion of practitioners~\citep{ID131:BosuEtAl2015,ID156:BaumEtAl2016,ID161:KononenkoEtAl2016,ID365:BosuEtAl2017,ID398:SutherlandVenolia2009, ID427:MacLeodEtAl2018,ID599:BernhartGrechenig2013,ID645:BaumEtAl2017,ID704:AlamiEtAl2019,ID713:RamEtAl2018,ID770:EbertEtal2019,ID709:SpadiniEtAl2019,ID820:SantosNunes2018} aim to better understand the MCR adoption. The goal is to identify the process variations and how developers perform code review.

Six studies gathered perceptions of reviewing practices in proprietary settings.
To understand why code reviews are used, \citet{ID156:BaumEtAl2016} interviewed professionals to identify factors influencing the adoption of MCR in the industry, reporting several variations observed in the review process. They also presented a list of factors shaping that process, which are organized into five categories: culture, development team, product, development process, and tools. The authors also reported that code review is most likely to remain in use when embedded into the development process. 
In a posterior study, \citet{ID645:BaumEtAl2017} refined this finding surveying professionals from commercial teams. They concluded that MCR (i.e.\ change-based code review) is the most common type of review and the risk of review fade away increases when there are no rules or conventions, which supports previous findings.
In addition to these findings, \citet{ID599:BernhartGrechenig2013} reported a positive effect of continuous review practice on the understandability and collective ownership in a particular project. 
Two works focused on investigating communication in MCR~\citep{ID398:SutherlandVenolia2009, ID131:BosuEtAl2015}. There is evidence of face-to-face and digital communication, not restricted to a tool that is dedicated to support MCR~\citep{ID398:SutherlandVenolia2009}. Overall, this communication is perceived as useful and helps developers understand the design rationale. Specifically, the review feedback is useful when it helps the author improve the code quality and triggers a code change~\citep{ID131:BosuEtAl2015}. Nevertheless, review comments with questions for clarifications are not considered useful by authors, but they may be useful to the reviewers.
Moreover, another study~\citep{ID820:SantosNunes2018} explored the opinion of professionals of how different factors (such as the patch size) influence internal outcomes of MCR (such as its duration and provided feedback). According to the study participants, the large number of active reviewers has a positive influence on the number of comments, while a large patch size and having more teams involved might have a negative influence on the review duration.

The perception of the practitioners of reviewing practices was also investigated in OSS, to understand why code review works in this context. \citet{ID704:AlamiEtAl2019} identified that in OSS the practice is more related to human and social aspects, particularly the hacker ethics. This study also observed that rejections and negative feedbacks are common in such a context, but there are professionals who perceive it as an opportunity for learning and technical growth due to the iterative improvement cycle. Concerning the practice quality, review feedback is also pointed out as a key aspect by~\citet{ID161:KononenkoEtAl2016}. The thoroughness of the feedback is associated with the review quality by practitioners as well as the reviewer's familiarity with the code, which is also observed in studies relying on objective data (discussed in later sections).

Lastly, the perception of practitioners on code review are also explored by research work  that consider both contexts, i.e.\ OSS and industry. \citet{ID770:EbertEtal2019} focused on exploring the main causes of confusion during a review and found as frequent reasons the missing rationale, discussion of non-funcional requirements, and lack of familiarity with code. To cope with this confusion, 13 strategies were identified, such as information requests, improved familiarity with the existing code, and off-line discussions. \citet{ID713:RamEtAl2018}, in turn, identified what makes a change easier to review. They concluded that reviewability can be defined through several factors, such as change description and size. Moreover, \citet{ID709:SpadiniEtAl2019} explored the opinion of practitioners on  test-driven code review (TDR), a variation of MCR when test-code is reviewed before the production code. They examined the perception of both problems and advantages of TDR, concluding that  developers prefer to review production code as they consider it more critical and tests should follow from it.

\begin{framed}
\noindent\textbf{Finding 2}: Multiple factors influence the adoption of MCR. In the industry, its adoption is influenced by culture, product, tools, development process, and team. In open-source projects, the influence is more related to human and social aspects, particularly hacker ethics.
\end{framed}

\begin{framed}
\noindent\textbf{Finding 3}: The review feedback is perceived as valuable when it provides an opportunity to learn or improve the code.
\end{framed}

\paragraph{\textbf{Social Aspects}}

MCR is a collaborative activity that relies on intensive human interaction. This motivated the researchers to investigate the social issues that emerge from the interaction among peers. 
There are three studies that focus on social aspects associated with MCR based on the developers' perceptions. \citet{ID125:GermanEtAl2018} investigate \emph{fairness}, analyzing how practitioners perceive the treatment given and received in code review. The results indicate that a significant proportion of participants perceives unfairness in MCR. This observation is more common among authors than reviewers. The other two studies~\citep{ID365:BosuEtAl2017,ID509:BosuCarver2013} explore the \emph{impression} of reviewers about their \emph{teammates}, more specifically, how it is formed and its impact on the practice. A key finding in both these studies is that MCR might impact on the impression formation, especially in building a perception of expertise. However, a poorly made code change may negatively impact this impression, also affecting how reviewers treat particular authors of changes in future reviews.

\paragraph{\textbf{Expected Benefits}}
The \emph{key} benefit expected of software inspection was the early defect identification. Though the analyzed studies also reported this as an expected benefit of MCR, only one survey~\citep{ID105:BacchelliBird2013} points out defect identification as the \emph{primary} expected benefit of MCR, being other benefits higher ranked. Most surveys report code improvement as the main desired benefit of MCR~\citep{ID156:BaumEtAl2016,ID365:BosuEtAl2017,ID427:MacLeodEtAl2018}, which means obtaining a better internal code quality, readability, and maintainability. Knowledge sharing and learning also emerged as key expectations~\cite{ID127:SadowskiEtAl2018,ID156:BaumEtAl2016,ID365:BosuEtAl2017}. In this case, people involved in a review desire to gain knowledge about the code, module, or coding style, for example. Moreover, practitioners also expect to discuss alternative solutions during the reviewing process, collaboratively developing new and better ideas~\cite{ID105:BacchelliBird2013,ID156:BaumEtAl2016}. 
In summary, (1) the two main expected benefits are \emph{code quality improvement} and \emph{defect identification}; and (2) MCR promotes additional benefits, such as \emph{knowledge sharing} and \emph{learning}.

\paragraph{\textbf{Challenges and Difficulties}}

Five studies~\citep{ID127:SadowskiEtAl2018,ID154:SpadiniEtAl2018,ID156:BaumEtAl2016,ID161:KononenkoEtAl2016,ID427:MacLeodEtAl2018} investigate what difficulties developers face when performing the code review activity, especially in an industrial environment. Understanding the motivation and purpose of a change has been pointed out as a challenge for practitioners of MCR~\citep{ID105:BacchelliBird2013, ID427:MacLeodEtAl2018,ID161:KononenkoEtAl2016,ID127:SadowskiEtAl2018}. There are studies~\citep{ID161:KononenkoEtAl2016,ID127:SadowskiEtAl2018} that indicate that gaining familiarity with the code is a technical challenge faced by reviewers because reviewing an unfamiliar code might result in misunderstandings. Moreover, in the context of reviewing a test code~\citep{ID154:SpadiniEtAl2018}, test files require developers to understand not only the code being reviewed (test files) but also the associated production files.

The other MCR challenges reported in the reviewed literature are more scattered. Two of the five studies~\citep{ID161:KononenkoEtAl2016, ID427:MacLeodEtAl2018} indicated time management as a challenge. In the study that focused on reviewing test code~\citep{ID154:SpadiniEtAl2018}, developers indicated that they have a limited amount of time to spend on reviewing, being driven by management policies to review production code instead of test code, which imposes a difficulty to review this type of code. Tools are also reported as challenges in two studies~\citep{ID127:SadowskiEtAl2018, ID161:KononenkoEtAl2016} because they are not suitable for a particular context or its customization may lead to misunderstanding.

In summary, \emph{code comprehension} has been the main challenge faced by developers when reviewing a code change. Other difficulties are also reported, such as time pressure and tool support.

\begin{framed}
\noindent\textbf{Finding 4}: The practitioners expect as effect of MCR the improvement of code quality and defect detection. Additionally, there are expectations related to knowledge sharing and learning. The main perceived challenge is to understand the code change, its purpose, and motivation.
\end{framed}
 
\subsection{Analysis of the Innerworkings of the MCR Process}

Six studies focus on understanding the internal mechanisms (i.e.\ innerworkings) of the MCR process, sharing findings, experiences and lessons learned from the adoption of MCR in real settings. Three of them share insights on the use of MCR in particular projects~\citep{ID158:BirdEtAl2015,ID147:Izquierdo-CortazarEtAl2017, ID594:BaysalEtAl2012}, while three studies report findings, lessons learned and recurrent practices of MCR in more than one setting~\citep{ID595:RigbyEtAl2012,ID116:RigbyBird2013,ID778:PaixaoMaia2019}.

Particular projects were investigated to understand the effects of process changes or the adoption of a particular tool on MCR, resulting in scattered findings. \citet{ID147:Izquierdo-CortazarEtAl2017} worked together with the developers of the Xen Project in the analysis of internal aspects of MCR to understand how the performance of code review evolves. In another study~\citep{ID594:BaysalEtAl2012}, data of Mozilla Firefox was examined to verify the differences between pre- and post-rapid release development, assuming that this change affects how the code is reviewed. Finally, \citet{ID158:BirdEtAl2015} described the developement and use of CodeFlow Analytics (CFA), a Microsoft's internal platform, which allows developers to explore the historical data of code review. Despite the specific findings of each study, they contribute with insights from real environments. The work targeting the Xen Project contributes with the approach used for analysis, while the report of CFA provides evidence of the positive impact of a tool for code review analytics. Moreover, there is evidence that post-rapid release is a successful approach to reduce the time patches wait for review, mainly considering patches from casual contributors, which are more likely to be abandoned.

In another study focusing on specific aspects of MCR, \citet{ID778:PaixaoMaia2019} analyzed data from multiple open-source systems. In this case, the study focuses on rebasing operations and their relationship with code review. As key contributions, the authors observed that rebasing occurs in most of reviews and also tend to tamper with the reviewing process, which might negatively affect the practice.

Lastly, two studies discuss more general insights. \citet{ID595:RigbyEtAl2012} introduced various lessons learned and recommendations based on the experience of code review in OSS that could be transferred to proprietary projects. They point out as lessons learned: (i) asynchronous, frequent and incremental reviews; (ii) invested experience reviewers; and (iii) empowerment of expert reviewers. They advocate the use of lightweight review tools and nonintrusive metrics, and the implementation of a review process. In another work~\citep{ID116:RigbyBird2013}, the analysis of several case studies led to the identification of common best practices on the use of MCR also in OSS. These practices indicated that code review (i) is a lightweight, flexible process; (ii) happens quickly, frequently, and early; (iii) change sizes are small; (iv) usually involves two reviewers; and (v) has changed from defect identification to a group problem-solving activity, in which reviewers prefer discussion and fixing code than reporting defects. Tool-supported review is suggested to provide the benefit of traceability, and its increased adoption is an indicator of success. 

\subsection{Cross-patch Analysis}

We identified in our set of \textsc{foundational studies} research work~\citep{ID712:HiraoEtAl2019,ID758:UedaEtAl2018,ID767:UedaEtAl2019} that explore patterns that occur across different patches of a project.
\citet{ID712:HiraoEtAl2019} investigated the impact of review linkage on MCR analytics. They extracted review linkage graphs of six projects and found that linkage rates range from 3\% to 25\%. They identified 16 types of review links, which are distributed into five categories: patch dependency, broader context, alternative solution, version control issues, and feedback related. Moreover, there is evidence that exploiting review linkage can improve the performance of reviewer recommendation approaches.
Differently, two studies aimed to identify behavioral patterns in the review feedback. \citet{ID758:UedaEtAl2018} compared how authors handle issues raised by automated checkers and manually indicated by reviewers. As a result, they found that authors repeatedly introduce the same types of problems despite the reviewer feedback, which does not occur with issues found by checkers. Then, \citet{ID767:UedaEtAl2019} explored  source code improvement patterns. They identified several patterns and grouped the eight most frequent into three categories, namely project-specific, readability-improvement, and language-specific. In these studies, the researchers aimed to reduce the cost of review, identifying issues that might be addressed by authors before code review.

\subsection{Analysis of MCR Outcomes}

The majority of \textsc{foundational studies} focus on objective and quantitative data from review practice, being this the most common type of study on MCR. These studies analyze the \emph{outcomes} of code review and how they are influenced by characteristics of a particular review, i.e.\ \emph{influence factors}. We split MCR outcomes into two groups: (i) \emph{internal outcomes}, which are associated with the MCR process (e.g.\ number of reviewers and number of comments); and (ii) \emph{external outcomes}, which are observed in the software product (code quality and defects). 

\begin{figure*}
 \centering
 \includegraphics[trim=0 60 0 20,clip,width=\linewidth]{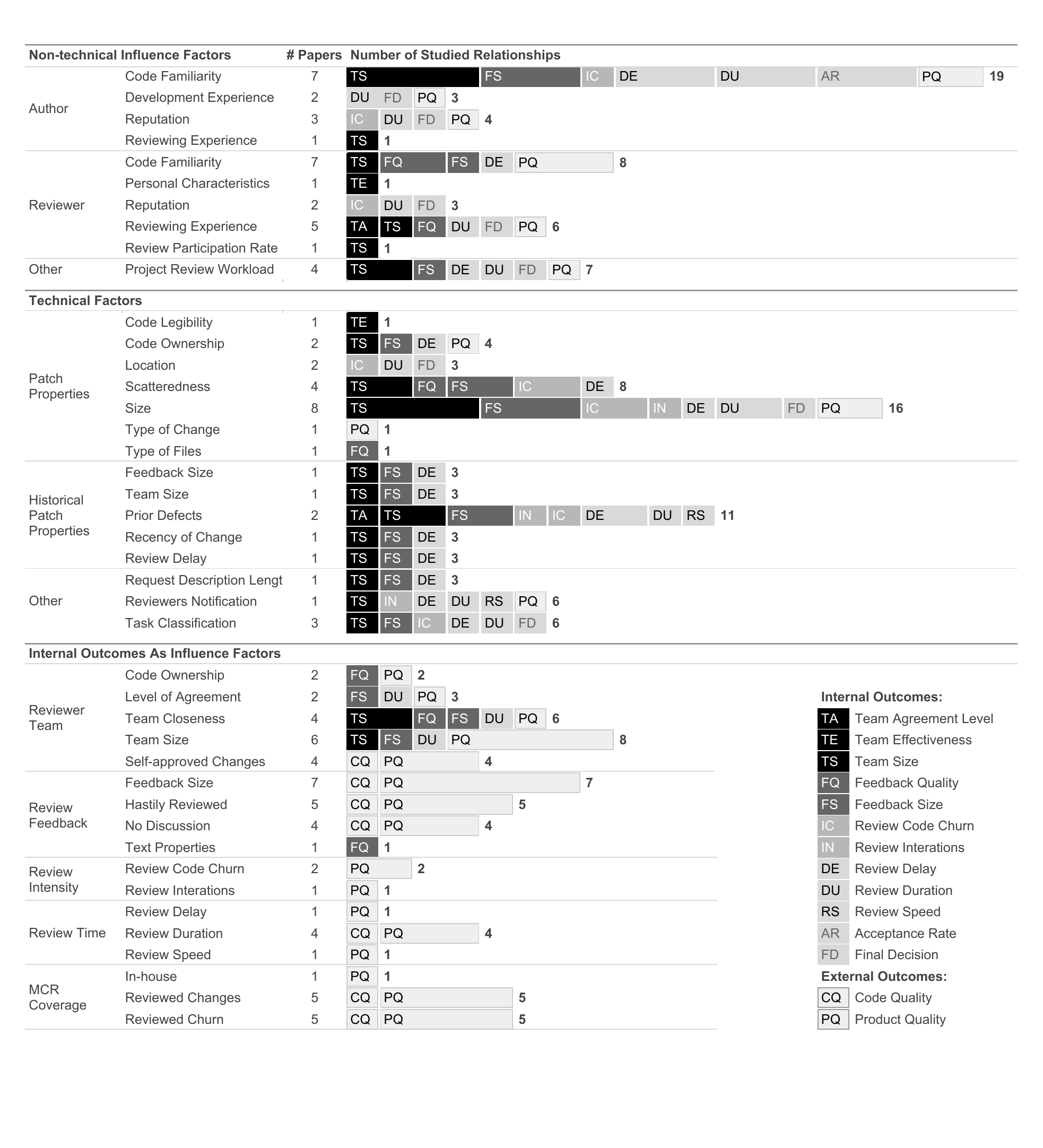}
 \caption{Studies that analyzed the relationship between influence factors and outcomes. On the left-hand side, there is the number of published studies that investigated the effects of an influence factor. On the right-hand side, there is the number of studies that investigated the effects of an influence factor over a particular outcome. A study can investigate the effect of an influence factor over more than one outcome.}
 \label{fig:IFvsO}
\end{figure*}

From the 48 works that focus on the analysis of MCR outcomes, we identified 27 studies exploring the correlation between 42 influence factors and 14 outcomes. We overview the studies of the relationship between influence factors and outcomes in Figure~\ref{fig:IFvsO}. On the left-hand side, there are influence factors. On the right-hand side, there is the number of times that the influence of a factor over an MCR outcome was investigated. Table~\ref{tab:influenceFactors} describes both non-technical and technical influence factors, along with the studies that explored their influence. Additionally, there are studies that investigate the influence of internal outcomes over external outcomes. These internal outcomes are detailed in Table~\ref{tab:influenceFactorsIO}.

The majority of the studies on MCR outcomes used data from OSS repositories, while five~\citep{ID104:ShimagakiEtAl2016,ID105:BacchelliBird2013, ID131:BosuEtAl2015, ID207:RahmanEtAl2017, ID820:SantosNunes2018} of them collected information from commercial projects and two~\citep{ID107:AlbayrakDavenport2010, ID624:MurakamiEtAl2017} from experiments. The results of these studies are discussed as follows.

\begin{table*}[t]
\caption{Influence factors analyzed by the \textsc{foundational studies}.}
\label{tab:influenceFactors}
\centering
\footnotesize
\setlength{\tabcolsep}{2pt}
\renewcommand{\arraystretch}{1.1}
\begin{tabularx}{\linewidth}{p{1.6cm} | p{3.4cm} | X | l}
\toprule
\textbf{Group} & \textbf{Influence Factor} & \textbf{Description} & \textbf{Studies} \\ 
\midrule
\multicolumn{4}{c}{\textbf{Non-Technical Factors}}\\ 
\hline
\multirow{3}{*}{Author}  
    & Code Familiarity         & Number of contributions authored by a developer in the project & \citep{ID191:KovalenkoBacchelli2018,ID206:BosuEtAl2014,ID318:LeeCarver2017,ID171:BosuCarver2014,ID644:ThongtanunamEtAl2017,ID480:KononenkoEtAl2015,ID816:RuangwanEtAl2019}\\ 
    & Development Experience & Number of contributions authored by a developer in general & \citep{ID653:BaysalEtAl2016, ID480:KononenkoEtAl2015}\\
    & Reputation             & Technical characteristics of the developer, such as working company, who authored the code & \citep{ID110:BellerEtAl2014, ID653:BaysalEtAl2016, ID206:BosuEtAl2014}\\ 
    & Reviewing Experience     & Number of completed reviews of code changes & \citep{ID816:RuangwanEtAl2019}\\
\hline
\multirow{4}{*}{Reviewer} & Code Familiarity & Number of review contributions made by a developer in the project & \citep{ID131:BosuEtAl2015, ID644:ThongtanunamEtAl2017, ID480:KononenkoEtAl2015, ID207:RahmanEtAl2017, ID331:MeneelyEtAl2014, ID651:McIntoshEtAl2016,ID816:RuangwanEtAl2019} \\
    & Personal Characteristics & Characteristics of the reviewer, such as age & \citep{ID624:MurakamiEtAl2017}\\
    & Reputation               & Technical characteristics of the developer, such as working company, who reviewed the code & \citep{ID110:BellerEtAl2014, ID653:BaysalEtAl2016} \\
    & Reviewing Experience     & Number of completed reviews of code changes & \citep{ID656:HiraoEtAl2016, ID207:RahmanEtAl2017, ID653:BaysalEtAl2016, ID480:KononenkoEtAl2015,ID816:RuangwanEtAl2019}\\ 
    & Review Participation Rate & Number of review invitations responded & \citep{ID816:RuangwanEtAl2019}\\
\hline    
Other & Project's Review Workload & Number of the review requests submitted to the code review tool in a period & \citep{ID644:ThongtanunamEtAl2017,ID653:BaysalEtAl2016, ID480:KononenkoEtAl2015,ID816:RuangwanEtAl2019}\\
\midrule
\multicolumn{4}{c}{\textbf{Technical Factors}}\\ 
\hline
\multirow{7}{*}{\parbox{1.4cm}{Patch Properties}}
      & Code Legibility & Presence of poor programming practices that affect code legibility or maintainability & \citep{ID107:AlbayrakDavenport2010}\\
    & Code Ownership  & Number of developers who submitted patches that impact the same files as the patch under review & \citep{ID651:McIntoshEtAl2016,ID644:ThongtanunamEtAl2017}\\ 
    & Location        & Location in the code change, such as the module it belongs to  & \citep{ID110:BellerEtAl2014, ID653:BaysalEtAl2016}\\ 
    & Scatteredness   &  Measure of the dispersion of the change, such as the number of files or directories in a review request &\citep{ID511:LiangMizuno2011, ID131:BosuEtAl2015, ID110:BellerEtAl2014, ID644:ThongtanunamEtAl2017}\\ 
    & Size            & Number of added or modified lines of code under review &\citep{ID820:SantosNunes2018, ID511:LiangMizuno2011, ID644:ThongtanunamEtAl2017, ID110:BellerEtAl2014, ID653:BaysalEtAl2016, ID206:BosuEtAl2014, ID331:MeneelyEtAl2014,ID816:RuangwanEtAl2019}\\ 
    & Type of Changes & Indication of the new or modified files  &\citep{ID206:BosuEtAl2014}\\ 
    & Type of Files   & Indication of the type of the file, such as source code or scripts &\citep{ID131:BosuEtAl2015}\\ 
\hline
\multirow{5}{*}{\parbox{1.4cm}{Historical Patch Properties}}   
	& Feedback Size         & Number of messages that were posted in the reviews of prior patches that impact the same files as the patch under review &\citep{ID644:ThongtanunamEtAl2017}\\ 
    & Number of Reviewers & Number of reviewers who provided feedback in the reviews of prior patches that impact same files as the patch under review &\citep{ID644:ThongtanunamEtAl2017}\\ 
     & Prior Defects       & Number of prior bug-fixing patches that impact the same files as the patch under review &\citep{ID644:ThongtanunamEtAl2017, ID172:ThongtanunamEtAl2015}\\ 
     & Recency             & Number of days since the last modification of the files &\citep{ID644:ThongtanunamEtAl2017}\\ 
     & Review Delay        &  Feedback delays of the reviewers of prior patches received &\citep{ID644:ThongtanunamEtAl2017}\\ 
 \hline    
\multirow{3}{*}{Other}  
    & Request Description Length & Number of words an author uses to describe a code change in a review request &\citep{ID644:ThongtanunamEtAl2017}\\ 
    & Reviewers Notification   & Indication of code review using broadcast (visible for all) or unicast (visible for a specific group) communication technology &\citep{ID450:ArmstrongEtAk2017}\\
    & Task Classification      & Indication of the change type based on purpose or priority, such as high-level priority bug fixing &\citep{ID644:ThongtanunamEtAl2017, ID110:BellerEtAl2014, ID653:BaysalEtAl2016}\\ 
\bottomrule
\end{tabularx}
\end{table*}
   
\begin{table*}
\caption{Internal outcomes as influence factors analyzed by the \textsc{foundational studies}.}
\label{tab:influenceFactorsIO}
\centering
\footnotesize
\setlength{\tabcolsep}{2pt}
\renewcommand{\arraystretch}{1.1}
\begin{tabularx}{\linewidth}{p{1.2cm} | p{2.5cm} | X | l}
\toprule
\textbf{Group} & \textbf{Influence Factor} & \textbf{Description} & \textbf{Studies} \\ 
\midrule
\multirow{5}{*}{\parbox{1.5cm}{Reviewer Team}}
    & Code Ownership      & Number of developers who uploaded a revision for the proposed changes &  \citep{ID207:RahmanEtAl2017, ID172:ThongtanunamEtAl2015}\\ 
    & Level of Agreement  & The proportion of reviewers that disagreed with the review conclusions & \citep{ID656:HiraoEtAl2016, ID172:ThongtanunamEtAl2015}\\ 
    & Team Closeness      & Number of distinct geographically distributed development sites or number of distinct teams associated with the author and reviewers & \citep{ID820:SantosNunes2018, ID131:BosuEtAl2015, ID331:MeneelyEtAl2014} \\
    & Team Size           & Number of reviewers that participate in the reviewing process & \citep{ID820:SantosNunes2018, ID134:ThompsonWagner2017, ID172:ThongtanunamEtAl2015, ID480:KononenkoEtAl2015, ID484:BavotaRusso2015, ID331:MeneelyEtAl2014}\\ 
    & Self-approved Changes & The proportion of changes approved for integration only by the original author & \citep{ID590:MoralesEtAl2015, ID103:McIntoshEtAl2014, ID104:ShimagakiEtAl2016, ID651:McIntoshEtAl2016}\\ 
\hline    
\multirow{4}{*}{\parbox{1.5cm}{Review Feedback}} 
    & Feedback Size       & Number of general comments and inline comments written by reviewers &  \citep{ID590:MoralesEtAl2015, ID104:ShimagakiEtAl2016, ID134:ThompsonWagner2017, ID172:ThongtanunamEtAl2015, ID484:BavotaRusso2015, ID480:KononenkoEtAl2015, ID651:McIntoshEtAl2016}\\ 
    & Hastily Review    & Number of hastily reviewed commits (changes approved for integration at a rate that is faster than 200 LOC/hour) & \citep{ID590:MoralesEtAl2015, ID103:McIntoshEtAl2014, ID104:ShimagakiEtAl2016, ID331:MeneelyEtAl2014, ID651:McIntoshEtAl2016}\\ 
    & No Discussion       & Number of accepted review requests without any review comments & \citep{ID590:MoralesEtAl2015, ID104:ShimagakiEtAl2016, ID103:McIntoshEtAl2014, ID651:McIntoshEtAl2016}\\  
    & Text Properties & Measure of textual features of review comments, such as stop word ratio & \citep{ID207:RahmanEtAl2017}\\  
\hline    
\multirow{2}{*}{\parbox{1.3cm}{Review Intensity}}  
    & Code Churn  & Number of lines added and deleted between revisions & \citep{ID172:ThongtanunamEtAl2015, ID104:ShimagakiEtAl2016}\\ 
    & Interations & Number of review iterations of a review request prior to its conclusion & \citep{ID172:ThongtanunamEtAl2015}\\ 
\hline    
\multirow{3}{*}{\parbox{1.3cm}{Review Time}}
    & Review Delay    & Time from the first review request submission to the first reviewer feedback& \citep{ID172:ThongtanunamEtAl2015}\\ 
    & Review Duration & Time from the first review request submission to the review conclusion & \citep{ID590:MoralesEtAl2015, ID172:ThongtanunamEtAl2015, ID104:ShimagakiEtAl2016, ID651:McIntoshEtAl2016}\\ 
    & Review Speed    & Rate of lines of code by an hour of a review request & \citep{ID172:ThongtanunamEtAl2015}\\ 
\hline    
\multirow{3}{*}{\parbox{1.3cm}{MCR Coverage}}
    & In-house         & Ratio of internal contributions in the project & \citep{ID104:ShimagakiEtAl2016}\\ 
    & Reviewed Changes & The proportion of committed changes associated with code reviews & \citep{ID590:MoralesEtAl2015, ID103:McIntoshEtAl2014, ID104:ShimagakiEtAl2016, ID134:ThompsonWagner2017, ID651:McIntoshEtAl2016}\\ 
    & Reviewed Churn   & The proportion of code churn reviewed in the past & \citep{ID590:MoralesEtAl2015, ID103:McIntoshEtAl2014, ID104:ShimagakiEtAl2016, ID134:ThompsonWagner2017, ID651:McIntoshEtAl2016}\\ 
\bottomrule
\end{tabularx}
\end{table*}

\begin{framed}
\noindent\textbf{Finding 5}: Most of the foundational knowledge on outcomes of MCR is based on data from open-source projects. From 48 studies that focus on analyzing internal or external outcomes of MCR,  41 (85.4\%) used data from OSS repositories. There are few studies in the industry (10.4\%) and experiments (4.2\%).
\end{framed}

\subsubsection{Internal Outcomes}

Internal outcomes of code review are those associated with characteristics of the MCR process. Examples of such characteristics are the number of involved reviewers, the number of provided comments and the amount of time that reviewers took to reach a decision about a review request.
Some studies are limited to the characterization of internal outcomes of specific projects, assessing and analyzing the values of these outcomes. Others inspect the relationship between influence factors and outcomes. From the 39 works that focus on internal outcomes, 33 used data from OSS projects, while the remaining ones used data from proprietary projects~\citep{ID105:BacchelliBird2013,ID131:BosuEtAl2015,ID820:SantosNunes2018,ID207:RahmanEtAl2017}. Moreover, two of the studies collected and used data produced by participants of an experiment~\citep{ID207:RahmanEtAl2017, ID624:MurakamiEtAl2017}. 
We grouped the investigated internal outcomes into five groups: \emph{reviewer team}, \emph{review feedback}, \emph{review intensity}, \emph{review time}, and \emph{review decision}. We next discuss the findings of each group.

\paragraph{\textbf{Reviewer Team}}

There are 14 studies that investigate the properties associated with the team of reviewers that are formed in code reviews. Few works~\citep{ID140:YangEtAl2017,ID149:BosuCarver2012,ID410:AsundiJayant2007} collected objective data with the single purpose of characterizing the participation of reviewers. The other remaining works examined the influence of different factors over outcomes related to reviewer teams. 

Most of the studies that analyze reviewer teams investigate the number of reviewers that engage in a code review, i.e.\ the \emph{team size}. In OSS projects, an average of one or two reviewers per request responded to a review request~\citep{ID140:YangEtAl2017,ID149:BosuCarver2012,ID410:AsundiJayant2007}. Also in the context of OSS, 16\%--66\% of the patches have at least one invited reviewer who did not respond~\citep{ID816:RuangwanEtAl2019}. As we discuss later, the participation of reviewers in code review share a positive relationship with product quality.

Eight studies~\citep{ID644:ThongtanunamEtAl2017, ID191:KovalenkoBacchelli2018, ID318:LeeCarver2017, ID820:SantosNunes2018, ID511:LiangMizuno2011, ID172:ThongtanunamEtAl2015, ID450:ArmstrongEtAk2017,ID816:RuangwanEtAl2019} explored factors influencing the team size. As key findings, these studies indicate that both the description length~\citep{ID644:ThongtanunamEtAl2017} and patch size~\citep{ID820:SantosNunes2018} influence the number of reviewers participating in a review---long patch descriptions and small patch size might increase the likelihood of attracting reviewers. Moreover, considering a proprietary project developed with distributed teams, \citet{ID820:SantosNunes2018} concluded that the team size decreases when more locations and teams are involved. \citet{ID816:RuangwanEtAl2019} also found that an experienced reviewer with higher review participation rate is more likely to respond a review invitation, i.e.\ a reviewer who has been actively responding to a review invitation in the past is more likely to respond to a new invitation. Another investigated aspect is the impact of familiarity with the project code on the team size. Three studies~\citep{ID191:KovalenkoBacchelli2018,ID318:LeeCarver2017,ID644:ThongtanunamEtAl2017} analyze the relationship between the number of previous contributions, i.e.\ changes of an author (as a means of measuring familiarity with the project) and the size of the team of reviewers. While two of the studies~\citep{ID191:KovalenkoBacchelli2018,ID644:ThongtanunamEtAl2017} did not find a significant relationship, \citet{ID318:LeeCarver2017} indicated a general trend that the number of active reviewers increases when the author's familiarity decreases, suggesting that newcomers receive more attention from invited reviewers. 

Going in another direction, \citet{ID140:YangEtAl2017} compared active and inactive reviewers of pull-requests. Their findings indicate that some super active reviewers lead code review, but inviting inactive reviewers would contribute to reducing the burden and speeding up the process. In fact, according to \citet{ID511:LiangMizuno2011}, authors prefer to invite more experienced reviewers, considering historical data, corroborating with the idea that there is a group of few reviewers that are overloaded in MCR.

In addition to team size, two other internal outcomes associated with reviewer teams have been explored, namely \emph{reviewer effectiveness} and \emph{reviewer agreement level}. The former was investigated in two empirical studies~\citep{ID107:AlbayrakDavenport2010, ID624:MurakamiEtAl2017}. One study~\citep{ID624:MurakamiEtAl2017} analyzed the effect of reviewer age on the efficiency and correctness of code review, but their findings did not provide evidence of a significant difference. The second study~\citep{ID107:AlbayrakDavenport2010}, in turn, examined how maintainability defects present in the code to be reviewed influences the effectiveness of reviewers, concluding that indentation issues have a negative impact on the reviewer performance.

The level of agreement among reviewers is investigated in two studies. \citet{ID172:ThongtanunamEtAl2015} analyze the relationship between a file with prior defects and the level of review disagreement, but the results do not show that there is a relationship between them. \citet{ID656:HiraoEtAl2016}, in turn, examined the influence of reviewers' reviewing experience on the frequency of their votes that disagreed with the review conclusions. The findings suggest that more experienced reviewers are more likely to have a higher level of agreement than less experienced reviewers. 

\begin{framed}
\noindent\textbf{Finding 6}: There is an average of one or two reviewers by review request in open-source projects. Overall, small code changes and long descriptions of review requests are more likely to attract reviewers. Developers prefer to invite experienced reviewers.
\end{framed}

\paragraph{\textbf{Review Feedback}} 

The most investigated internal outcome is review feedback, with a total of 26 papers. These studies explored the comments made by reviewers. We identified three examined aspects: (i) content information~\citep{ID105:BacchelliBird2013,ID154:SpadiniEtAl2018,ID202:EfstathiouSpinellis2018,ID206:BosuEtAl2014,ID532:BiaseEtAl2016,ID642:LiEtAl2017,ID465:EbertEtAl2017,ID770:EbertEtal2019,ID783:PaulEtAl2019,ID807:ElAsriEtAl2019,ID759:NorikaneEtAl2018,ID812:WangEtAl2019,ID705:PascarellaEtAl2018,ID710:ZanatyEtAl2018,ID746:EbertEtAl2018}; (ii) size, typically in terms of amount of comments~\citep{ID511:LiangMizuno2011,ID149:BosuCarver2012, ID644:ThongtanunamEtAl2017,ID318:LeeCarver2017,ID191:KovalenkoBacchelli2018, ID820:SantosNunes2018,ID172:ThongtanunamEtAl2015,ID656:HiraoEtAl2016,ID759:NorikaneEtAl2018,ID771:JiangEtAl2019}; and (iii) quality, in terms of, e.g., usefulness~\citep{ID207:RahmanEtAl2017,ID131:BosuEtAl2015}. 

To further understand what has been discussed within code reviews, there is research work that explored the nature of dialogs and the concerns raised by human reviewers. \citet{ID105:BacchelliBird2013} manually analyzed and classified MCR comments from Microsoft projects, creating categories for emerged themes. \citet{ID154:SpadiniEtAl2018} reproduced this analysis using comments from the test code review of OSS projects. Both studies identified code improvement, understanding, social communication, and defects as the most frequent discussion topics. Two studies~\citep{ID532:BiaseEtAl2016, ID206:BosuEtAl2014} examined security concerns raised in the review feedback, leading to identified categories related to the domain and language-specific issues~\citep{ID532:BiaseEtAl2016} and race conditions~\citep{ID206:BosuEtAl2014}. While the mentioned studies performed a manual analysis of the comments, \citet{ID642:LiEtAl2017} proposed a taxonomy of review comments on pull-requests and an automatic classifier based on that taxonomy. They used the classifier to identify the typical review patterns in OSS projects and found that most are about code correction and social interactions. 

\citet{ID710:ZanatyEtAl2018} investigated design-related discussions in code reviews, concluding that this aspect is not commonly discussed. However, when design issues are raised in the review feedback, they are considered constructive, offering alternative solutions. Moreover, two studies~\citep{ID705:PascarellaEtAl2018,ID746:EbertEtAl2018} specifically analyzed the questions and answers in the review feedback to identify the information that reviewers need and their communicative intentions. As contributions, both studies found that questions are used to ask an action of the author related to a suggestion of an alternative solution. The request of confirmation or clarification of the correct understanding were also topics found in these studies.

We also identified research work focused on confusion~\citep{ID465:EbertEtAl2017,ID770:EbertEtal2019} and sentiments~\citep{ID783:PaulEtAl2019,ID807:ElAsriEtAl2019} expressed by reviewers. With respect to confusion, \citet{ID465:EbertEtAl2017} initially studied the feasibility of analyzing the confusion using linguistic feature, they then complement this investigation by exploring the reasons for the confusion and on how developers cope with it~\citep{ID770:EbertEtal2019}. Several reasons were found for confusion, being most frequent the missing of rationale, discussion of non-functional requirements, and lack of familiarity with code. Concerning the topic of sentiments, \citet{ID807:ElAsriEtAl2019} conducted a broad empirical study, while \citet{ID783:PaulEtAl2019} focused on the differences in expressions between male and female developers during code review. These studies observed differences depending on the position in the collaboration network and the gender of reviewers. While peripheral contributors have more outliers in expressing positive and negative sentiments, the core developers are neutral when commenting a review~\citep{ID807:ElAsriEtAl2019}. The results also suggest that females are less likely to express sentiments than males~\citep{ID783:PaulEtAl2019}.

In addition to the discussed studies, two other works investigate specific aspects using the review feedback. \citet{ID812:WangEtAl2019} manually analyzed reviewed changes and classified the reasons for abandoning them in 12 categories. The top three categories are: (i) duplicate, in which are included changes that were abandoned because they were similar to others; (ii) lack of feedback, when authors did not respond the review feedback or when nobody responded the review request; and (iii) contributor operation, when occurs erroneous operations. Nevertheless, most changes are abandoned due to duplication. \citet{ID759:NorikaneEtAl2018}, in turn, analyzed the review feedback to understand what encourages a contributor to continue with an open-source project. As a result, the study observed similarities in the early contributions of those who become long-term contributors (LTC) and those who become short-time contributors (STC). In summary, those who become an LTC submitted more code changes and received more review feedback than STCs. Moreover, there are more similarities between the reasons for rejecting LTCs, while STCs received more scattered motives for rejection.

Focusing on the feedback size, two studies~\citep{ID149:BosuCarver2012,ID511:LiangMizuno2011} analyzed the amount of discussion by review in OSS projects and observed an average of two or three comments per review. When examining the review data of reopened pull-requests, \citet{ID771:JiangEtAl2019} found an average of two comments for non-reopened pull requests and seven comments for reopened requests. This low number, together with the limited reviewer participation in reviews, motivated the investigation of various factors influencing the feedback, which was done in six papers~\citep{ID191:KovalenkoBacchelli2018,ID318:LeeCarver2017,ID644:ThongtanunamEtAl2017,ID656:HiraoEtAl2016,ID820:SantosNunes2018,ID172:ThongtanunamEtAl2015}. 

Three studies~\citep{ID191:KovalenkoBacchelli2018,ID318:LeeCarver2017,ID644:ThongtanunamEtAl2017} examine the relationship between the familiarity with the project of the author submitting the change and the feedback size. Their key finding is that the number of comments increases as the familiarity decreases, indicating higher involvement of reviewers in review requests made by novices. \citet{ID656:HiraoEtAl2016}, in turn, provided evidence that a review request with a reviewer with a lower level of agreement is more likely to have a longer discussion length. Similarly to team size, the feedback size is also influenced by the patch size. \citet{ID820:SantosNunes2018} reported that the larger the patch, the lower the comment density, according to their analysis. Despite this, \citet{ID644:ThongtanunamEtAl2017} indicated that the more lines changed in the patch, the more likely the patch is discussed. In previous work, \citet{ID172:ThongtanunamEtAl2015} also found that risky files, i.e., files with prior defects, tend to undergo reviews that have shorter discussions and more revisions without reviewer feedback than normal files do.

Differently from the studies above, three studies observed the quality and technical aspects of the provided comments. \citet{ID131:BosuEtAl2015} and \citet{ID207:RahmanEtAl2017} explored factors that influence the usefulness of code review comments. The first work made this analysis at Microsoft, while the second explored the textual features and developer experience to ground the proposal of a usefulness predictor. Both studies found that the code familiarity of the reviewer influences the feedback quality. \citet{ID207:RahmanEtAl2017} also found some variation among textual properties between useful and non-useful comments. Motivated by these two studies~\citep{ID131:BosuEtAl2015,ID207:RahmanEtAl2017}, \citet{ID202:EfstathiouSpinellis2018} presented a preliminary investigation with OSS data to examine facets of language in the comments. They observed a collocation of source code and linguistic coherence in the review messages, suggesting that this might support future research on the analysis of usefulness in review comments.

\hfill \break \break \break

\begin{framed}
\noindent\textbf{Finding 7}: The most frequent discussion topics in the review feedback are related to code improvement, understanding, and social interactions. The reviewers also discuss alternative solutions related to design, and most of the suggestions are made using questions. Moreover, in the feedback, reviewers express confusion, often due to missing rationale, discussion of non-functional requirements, and the lack of familiarity with code.
\end{framed}

\begin{framed}
\noindent\textbf{Finding 8}: In open-source projects there is an average of two or three review comments per request. Newcomers tend to receive more attention. The reviewer more familiar with the code is more likely to provide useful feedback.
\end{framed}

\paragraph{\textbf{Review Intensity}}

We now focus on work that targets the review intensity, which refers to analyses of the number of iterations made during code reviews and the code churn (the delta between the submitted and accepted code). Seven papers fall into this category~\citep{ID110:BellerEtAl2014,ID171:BosuCarver2014,ID653:BaysalEtAl2016,ID511:LiangMizuno2011,ID172:ThongtanunamEtAl2015,ID450:ArmstrongEtAk2017,ID449:UedaEtAl2017}. Two~\citep{ID449:UedaEtAl2017, ID653:BaysalEtAl2016} investigate the content of the code churn, while those remaining make an analysis of the correlation between influence factors and the review intensity. 

The two studies on the content of changes made in the code under review explored different aspects. \citet{ID449:UedaEtAl2017} studied how the author of changes fixes \texttt{if} statements during the code review, identifying symbolic operators that are typically changed, such as parentheses. In contrast, \citet{ID110:BellerEtAl2014} explored the problem types fixed during code review, manually classifying the changes into a defect categorization, as well as analyzing what triggers them. The results of this work indicate a 75:25 ratio between evolvability and functional changes, being 10\%--22\% of these changes not triggered by review feedback. 

Considering code churn, the studies on review intensity suggest that \emph{technical} aspects potentially affect the delta between the submitted and accepted code. \citet{ID110:BellerEtAl2014} indicated that bug-fixing tasks lead to fewer changes, while patches with more altered files and lines of code lead to higher code churn. Similarly, \citet{ID511:LiangMizuno2011} explored technical aspects but did not find a strong correlation between patch content and churn. In another direction, \citet{ID171:BosuCarver2014} analyzed the reputation of the author of the code change, i.e.\ core and peripheral developer, and its relation with several factors, including the code churn. Despite several identified differences, the result of the analysis related to the number of patches per review requests is inconclusive. 

Investigating the number of review iterations, \citet{ID653:BaysalEtAl2016} found that larger changes have more rounds of revisions. \citet{ID172:ThongtanunamEtAl2015}, in turn, investigated review intensity in risky files, i.e.\ files with prior defects. Although their findings suggest that risky files tend to undergo reviews that have fewer iterations, the same analysis indicated that these files churn more during MCR. Finally, \citet{ID450:ArmstrongEtAk2017} reported that patches reviewed using unicast technology (when a review request is visible for a targeted group) undergo more iterations than those reviewed using broadcasts technologies (when all those subscribed to a medium can see the review request).

\begin{framed}
\noindent\textbf{Finding 9}: Most of the problems fixed during code review are related to evolvability and are triggered by review feedback. Small code changes and files with prior defects tend to undergo few rounds of code review.
\end{framed}

\paragraph{\textbf{Review Time}}

\emph{Time} is another aspect of the MCR process that has been investigated. Studies focus in particular on (i) review duration (the total amount of time taken to reach a decision), (ii) the first response delay, and (iii) review speed. In total, 14 papers targeted this topic. Four of them~\citep{ID140:YangEtAl2017,ID149:BosuCarver2012,ID464:ThongtanunamEtAl2015,ID657:KerzaziElAsri2016} characterize the review interval in particular contexts. The other papers explore the relationship between influence factors and review time aspects.

\citet{ID140:YangEtAl2017} aimed to understand why code review is considered a time-consuming process. They analyzed pull-requests from Rails, an open-source project, and found that more than 40\% of its pull-requests are closed in more than ten days, being considered a long time for reviewers to complete the review. \citet{ID149:BosuCarver2012} also assessed a typical review interval in OSS projects together with the delay for the code author to receive the first feedback. The median review interval is 3--4 days, while the first review feedback is received promptly in most cases. Furthermore, also in the OSS context, \citet{ID657:KerzaziElAsri2016} examined both technical and socio-technical interactions among contributors in code review. They identified behavioral patterns, reporting that core developers are more likely to have a shorter review interval than peripheral developers~\citep{ID657:KerzaziElAsri2016}. Lastly, \citet{ID464:ThongtanunamEtAl2015} investigated reviews with code-reviewer assignment problems and the impact in review duration. They analyzed the content of comments and found that: (i) 4\%--30\% of the reviews have the assignment problem; and (ii) these reviews require, on average, 12 days longer to approve code changes.

In addition to the discussed findings on review duration, many studies~\citep{ID820:SantosNunes2018,ID171:BosuCarver2014,ID172:ThongtanunamEtAl2015,ID191:KovalenkoBacchelli2018,ID318:LeeCarver2017,ID450:ArmstrongEtAk2017,ID653:BaysalEtAl2016,ID656:HiraoEtAl2016} explore which factors might be related to the total amount of time taken to reach a review decision. Although some did not identify a significant relationship, others provided evidence of factors influencing review duration. The key findings of these studies indicate that the review takes less time when (i) the authors of code changes are more experienced and familiar with the project~\citep{ID653:BaysalEtAl2016,ID171:BosuCarver2014,ID318:LeeCarver2017}; (ii) the involved reviewers have high reviewing experience~\citep{ID653:BaysalEtAl2016}; (iii) the review queue is short~\citep{ID653:BaysalEtAl2016}; (iv) the patch size is small~\citep{ID820:SantosNunes2018}; (v) there are few prior defects in the reviewed files~\citep{ID172:ThongtanunamEtAl2015}; (vi) there are few active reviewers, and they are from the same team and location~\citep{ID820:SantosNunes2018}; (vii) the historical level of agreement of involved reviewers is high~\citep{ID656:HiraoEtAl2016}; and (viii) the review feedback contains positive sentiments~\citep{ID807:ElAsriEtAl2019}. Moreover, \citet{ID771:JiangEtAl2019} analyzed characteristics of reopened pull-requests and their impact on code review, reporting that these requests take more review time to be concluded.

Other five studies~\citep{ID171:BosuCarver2014,ID172:ThongtanunamEtAl2015,ID191:KovalenkoBacchelli2018,ID450:ArmstrongEtAk2017,ID644:ThongtanunamEtAl2017} conducted a similar analysis of the factors influencing the first response delay of a review request. Similarly as above, some studies did not find evidence to support this relationship, while others reported both technical and non-technical aspects associated with this outcome. The key findings are that authors of code changes that are more familiar with the project receive faster first feedback~\citep{ID171:BosuCarver2014}. In contrast, if a patch has files with prior defects~\citep{ID172:ThongtanunamEtAl2015} and these files historically received a slow response~\citep{ID644:ThongtanunamEtAl2017}, then the first response also tends to be slower.

Another aspect of MCR that has been studied is the review speed (reviewed lines of codes per hour). \citet{ID172:ThongtanunamEtAl2015} investigated whether the speed varies depending on the presence of defects in a prior release, while \citet{ID450:ArmstrongEtAk2017} studied the difference in the review speed when using unicast or broadcast as communication technology. A key finding of these studies is that files with prior defects tend to have a faster review rate than the reviews of normal files. 

\begin{framed}
\noindent\textbf{Finding 10}: The review request takes less time to be completed when involving experienced authors familiarized with the code, experienced and active reviewers with a high level of agreement. Moreover, the review duration reduces when there is a short review queue, the code change is small, the files have few prior defects, and the review feedback contains positive sentiments.
\end{framed}

\paragraph{\textbf{Review Decision}}

The last group of influence factors refers to the result of code reviews (accept or reject), that is, the review conclusion. This is one of the least explored topics with nine identified papers~\citep{ID653:BaysalEtAl2016, ID171:BosuCarver2014, ID318:LeeCarver2017, ID191:KovalenkoBacchelli2018, ID149:BosuCarver2012, ID206:BosuEtAl2014, ID255:HiraoEtAl2015,ID771:JiangEtAl2019,ID807:ElAsriEtAl2019}. The observed findings are scattered.

\citet{ID255:HiraoEtAl2015} investigated how many review requests followed the simple majority method of voting to decide on the acceptance or rejection of code changes. With a case study in an OSS project, the researchers aimed to understand the criteria for integrating a changeset. Their results indicate that only 59.5\% of the requests followed the simple majority method and requests with more negative votes than positive votes were likely to be rejected.
\citet{ID149:BosuCarver2012}, based on the examination of the proportion of review requests rejected in Asterisk and MusicBrainz (OSS projects), identified that 7.5\% and less than 1\% of the requests are not accepted, respectively. \citet{ID206:BosuEtAl2014}, in contrast, focused on uncovering the changes containing which types of vulnerabilities are more likely to be abandoned. They concluded that MCR leads to the identification of common types of vulnerabilities. Moreover, code is more likely to be vulnerable when it is authored by less experienced contributors, has a high number of lines changed, and consists of modified (as opposed to new) files.

Lastly, there are few studies exploring the factors influencing the review outcome and acceptance rate. \citet{ID653:BaysalEtAl2016} analyzed multiple factors and concluded that the review outcome is most affected by author development experience. The findings of \citet{ID807:ElAsriEtAl2019} also indicate that negative review comments share a relationship with the likehood of an unsuccessful review. Similarly to \citet{ID653:BaysalEtAl2016}, three studies~\citep{ID171:BosuCarver2014, ID191:KovalenkoBacchelli2018, ID318:LeeCarver2017} focus on the the author's familiarity with the project, with results indicating that the acceptance rate is lower for newcomers. In addition, \citet{ID771:JiangEtAl2019} found that reopened pull requests have lower acceptance rates in the code review.

\begin{framed}
\noindent\textbf{Finding 11}: The accepted code changes are usually authored by experienced developers and contain review feedback with positive sentiments. Considering the review feedback analysis, the main reasons to abandon a code change are duplication and lack of feedback.
\end{framed}

\subsubsection{External Outcomes}

The results discussed in the previous section consists of the analysis of how different factors influence outcomes associated with the code review process. However, these are not properties that are externally perceived (e.g., how intense the discussion is). We now discuss work that focus on external outcomes, which are mainly improvement of code quality (design and programming practices) and product quality (reduction of defects).
We identified 13 papers that explored quantitative data collected from repositories, not only with code review data but also the code being reviewed. From these papers, only one~\citep{ID590:MoralesEtAl2015} focused on code quality, while the others~\citep{ID206:BosuEtAl2014,ID480:KononenkoEtAl2015,ID331:MeneelyEtAl2014,ID651:McIntoshEtAl2016,ID104:ShimagakiEtAl2016,ID450:ArmstrongEtAk2017,ID172:ThongtanunamEtAl2015,ID134:ThompsonWagner2017,ID484:BavotaRusso2015,ID103:McIntoshEtAl2014,ID532:BiaseEtAl2016,ID750:AnEtAl2018} on product quality. 

\citet{ID590:MoralesEtAl2015} studied the impact of MCR on software design by examining how the incidence of seven anti-patterns is affected by the review coverage and participation. Their findings indicate that components with low coverage or low review participation are more likely to have occurrences of anti-patterns, but with variances observed across the analyzed projects.

Focusing on a different perspective, four other studies~\citep{ID206:BosuEtAl2014,ID331:MeneelyEtAl2014,ID134:ThompsonWagner2017,ID532:BiaseEtAl2016} analyze OSS projects to investigate factors influencing security aspects of code that went through code review. \citet{ID532:BiaseEtAl2016} presented a case study of security aspects of Chromium, analyzing several aspects of code review data. With respect to the MCR coverage, the researchers reported that code reviews tend mostly to miss language-specific and domain-specific issues, such as buffer overflows and Cross-Site Scripting. The other three studies then provide evidence of what might increase the likelihood of a code change to contain a security flaw after being checked in code review. As key findings, the mentioned works indicate that (i) the majority of the vulnerable code is written by the most experienced authors, although the less experienced authors' changes are 1.5 to 24 times more likely to be vulnerable~\citep{ID206:BosuEtAl2014}; (ii) more lines churned increased the probability of a patch to contain a vulnerability~\citep{ID206:BosuEtAl2014,ID331:MeneelyEtAl2014}; (iii) modified files are more likely to have vulnerabilities than new files~\citep{ID206:BosuEtAl2014}; (iv) vulnerable files tend to have more involved reviewers, with lower security-experience~\citep{ID331:MeneelyEtAl2014}. Furthermore, the study of \citet{ID134:ThompsonWagner2017} reports that code review appears to reduce the number of issues and security issues, revealing that there is relationship between review coverage (assessed by unreviewed pull requests and unreviewed churn), and review participation (measured by average commenters, mean discussion comments, and mean review comments). 

In addition to the studies on security aspects, the remaining papers related to external outcomes are focused on the overall software quality. \citet{ID484:BavotaRusso2015} identified that unreviewed code changes have over two times more chances of inducing bugs than reviewed changes. They also reported that there is a difference between the quality attributes complexity and readability of code components in reviewed and unreviewed commits. Despite this positive impact of reviewing practices on software quality, \citet{ID480:KononenkoEtAl2015} identified that 54\%--56\% of code reviews missed bugs. 

Considering factors influencing the bug proneness of reviewed code changes, the likelihood of post-release defects increased as the involved reviewers have fewer reviewing experience~\citep{ID480:KononenkoEtAl2015}, and they are less familiar with the project (lack subject matter expertise)~\citep{ID651:McIntoshEtAl2016}. Review queue also has an impact on whether reviewers catch bugs, being longer review queues more related to defect-proneness changes~\citep{ID480:KononenkoEtAl2015}. Analyzing review practices, \citet{ID172:ThongtanunamEtAl2015} identified that future-defective files are less intensely scrutinized, having less participation of reviewers, and a faster rate of code checking than files without post-release defects. Moreover, both \citet{ID172:ThongtanunamEtAl2015} and \citet{ID480:KononenkoEtAl2015} found a relationship between a small number of reviewers and the increasing likelihood of missing issues. This finding is in contrast with the result in the work of \citet{ID331:MeneelyEtAl2014}, which shows that vulnerable files tend to have more involved reviewers. Additionally, \citet{ID450:ArmstrongEtAk2017} observed that review using unicast communication technology has fewer defects than broadcast communication. 

\citet{ID750:AnEtAl2018} focused on crashed reviewed code, a type of defect with severe implications. The goal is to understand why reviewed change still crashes. As result, they found that those reviewed changes are usually related to performance, refactoring, fix previous crashes, and new functionality. Moreover, they found that the crashses are mainly motivated by memory and semantic errors, which were not detected during code review.

Finally, \citet{ID103:McIntoshEtAl2014} and \citet{ID104:ShimagakiEtAl2016} studied post-release defects and their relationship with code review coverage and review participation. While \citet{ID103:McIntoshEtAl2014} analyzed MCR practices in OSS projects, \citet{ID104:ShimagakiEtAl2016} replicated the study in a proprietary setting at Sony Mobile. Despite the slight differences in metrics for assessing review participation and coverage (metrics added in the more recent study), \citet{ID104:ShimagakiEtAl2016} reported that the relationship associated with software quality identified in the original research is not consistent with that identified in the proprietary setting. Despite the differences, both studies indicate that code review practices share a strong association with defect-proneness.

\begin{framed}
\noindent\textbf{Finding 12}: The product quality tends to increase when code changes are reviewed. More participation in code reviews, more experience with the reviewing activity, and reviewers more familiar with the project are also factors that reduce the likelihood of post-release defects.
\end{framed}

\subsection{Human Aspects of Reviewers}

Code review is mostly based on the subjective human evaluation of code changes and involves intensive human interactions. Therefore, some researchers explored the \emph{behavior} of reviewers during code review, how they check the changed code, and their willingness to participate in MCR. Nine studies go to this direction: (i) three~\citep{ID209:BegelVrzakova2018,ID291:UwanoEtAl2006,ID408:ChandrikaEtAl2017} rely on eye tracking; (ii) two~\citep{ID277:FloydEtAl2017,ID613:DuraesEtAl2016} rely  on functional magnetic resonance imaging (fMRI); and (iii) the remaining four~\citep{ID144:KitagawaEtAl2016,ID211:DunsmoreEtAl2000,ID815:BaumEtAl2019,ID709:SpadiniEtAl2019} analyze the reviewer behavior with simulation and experimental data.

Studies that used eye tracking~\citep{ID209:BegelVrzakova2018,ID291:UwanoEtAl2006,ID408:ChandrikaEtAl2017} aim to understand how reviewers check the code. \citet{ID291:UwanoEtAl2006} conducted an experiment with professionals to characterize the overall performance of individuals during code review. They found that the subjects are likely to read the whole lines briefly, then concentrate on particular sections. \citet{ID209:BegelVrzakova2018} are more specific and analyze the eye movements that triggers review comments, presenting a classification of five kinds of code elements that might act as a trigger. Finally, \citet{ID408:ChandrikaEtAl2017} investigated the eye-tracking trait differences of subjects with and without programming skills to understand visual attention. The authors concluded that the key aspect for MCR is attention span on error lines and comments and better code coverage. In summary, these three studies suggest that reviewers first examine all lines of changed code, then focus on specific proportions, which might be influenced by programming skills. 

Exploring fMRI, two studies~\citep{ID277:FloydEtAl2017,ID613:DuraesEtAl2016} aim to understand the brain activity of reviewers to identify patterns for analysis. One of the studies~\citep{ID277:FloydEtAl2017} was conducted with students in an attempt to relate tasks performed by individuals with patterns of brain activation. The authors compared tasks of code review, code comprehension, and English prose review (a snippet of English writing marked up with edits) and identified distinct neural representations. They also found that a programming language is treated more like a natural language when an individual has more expertise. The other research~\citep{ID613:DuraesEtAl2016} focuses on brain activity patterns when the reviewer identifies a bug. The researchers also found specific brain regions where activation increased during code review, specifically the areas associated with language processing and mathematics. They showed that particular brain activity patterns can be related to the decision-making moment of suspicion/bug detection.

Finally, the remaining studies investigate particular aspects of the review behavior. \citet{ID144:KitagawaEtAl2016} aimed to understand the reviewer participation in MCR using simulation. It consists of a model of a reviewing situation based on a snowdrift game. A key finding is that a reviewer cooperates with others when the benefit of a review is higher than its cost. The other study~\citep{ID211:DunsmoreEtAl2000} is an empirical investigation of defect detection in object-oriented (OO) programs, concluding that defects that require information spread throughout the software to be identified are hard to find and the OO code structure favors this type of defect. \citet{ID815:BaumEtAl2019} then experimentally analyzed the association between aspects of the cognitive load of reviewers and their performance. As key findings, they found a correlation between working memory capacity and the reviewer's effectiveness of finding delocalized defects, as well as evidence of the negative impact of larger and complex code changes on review performance. Finally, \citet{ID709:SpadiniEtAl2019} explored the influence of the order that the test code is presented to the reviewer, i.e.\ test then production code or production then test code. Based on a controlled experiment, they observed that reviewers in a test-first review find the same proportion of defects in production file and more defects in test code. 

\begin{framed}
 \noindent\textbf{Finding 13}: The programming experience might influence reviewers' attention and brain work during the code checking. Specific brain regions associated with language processing and mathematics have increased activity when reviewing a code, and experienced developers tend to perceive programming languages like a natural language.
\end{framed}

\subsection{Relationship with MCR}

The \textsc{foundational studies} discussed previously focus solely on the MCR practice. The last group of \textsc{foundational studies} consists of research that analyzed how MCR is related to other approaches within software development. We discuss identified papers in two groups. The first compares MCR with other verification techniques, and the second analyzes the impact of MCR in other practices of the software development.

\subsubsection{Comparison with Verification Techniques}

There are experiments that compare MCR with verification techniques, namely pair programming~\citep{ID077:Muller2005,ID560:SwamiduraiEtAl2014} and testing~\citep{ID487:RunesonAndrews2003}. They involved only students, and the overall goal is to examine the effectiveness of one technique with respect to another.

\citet{ID077:Muller2005} and \citet{ID560:SwamiduraiEtAl2014} compared code review and pair programming aiming to verify which has a higher impact in terms of cost. In both studies, the participants are divided into the ones using pair programming to execute a task, and those who work individually with the assistance of a code review phase. The difference between the two studies is that in Swamidurai et al.'s experiment~\citep{ID560:SwamiduraiEtAl2014} the techniques are adopted in the context of the Test-Driven Development (TDD) environment. As a key finding, \citet{ID077:Muller2005} indicated that pairs are as cheap as single developers if both are forced to produce code of similar correctness. In contrast, when taking into account programs of different levels of correctness, pairs provide code with fewer failures at a higher expense, although the difference is not statistically significant. Therefore, this study suggests that pair programming and individual review may be interchangeable in terms of cost~\citep{ID077:Muller2005}. However, in the context of TDD, \citet{ID560:SwamiduraiEtAl2014} found evidence that programs with similar quality can be produced using peer review with 28\% lower cost than using pair programming.

While the comparisons with pair programming focused on the cost, there is a study that compares testing practices and code review focuses on the capability of detecting defects. \citet{ID487:RunesonAndrews2003} compared unit testing with code review, investigating the detection and isolation of the underlying sources of the defects. As a result, they reported differences, being code review more effective in terms of time spend and isolation, and testing finds more failures~\citep{ID487:RunesonAndrews2003}. 

Considering the discussed findings, we highlight that some were published more than a decade ago (from 2003 to 2014) and, since then, the MCR key goal has shifted from defect detection to problem-solving~\citep{ID116:RigbyBird2013} and tool-support became popular~\citep{ID105:BacchelliBird2013}. The expected benefits of practitioners have also been changing, as discussed. In our SLR, we did not identify more recent comparisons of MCR with other verification techniques.

\begin{framed}
 \noindent\textbf{Finding 14}: MCR and pair programming are interchangeable in terms of cost, except when the latter is adopted within test-driven development---in this case, MCR has lower cost. Unit testing finds more failures than MCR, but the latter requires less time in the detection and isolation of the underlying sources of the defects.
\end{framed}

\subsubsection{Interaction with Development Practices}

The last group of \textsc{foundational} studies has five analyses involving MCR and its relationship with other software development practices. Three studies focus on quality, verifying its relationship with continuous integration~\citep{ID126:RahmanRoy2017,ID768:ZampettiEtAl2019} and static analysis~\citep{ID538:PanichellaEtAl2015}. One study investigates the intent and awareness of developers when performing changes concerned with architectural aspects~\citep{ID785:PaixaoEtAl2019}. Finally, the fifth study explores code review and its association with code ownership~\citep{ID267:ThongtanunamEtAl2016}. All these five works examine code review data from OSS projects.

The interplay between reviewing practices and continuous integration (CI) has been investigated in two studies~\citep{ID126:RahmanRoy2017,ID768:ZampettiEtAl2019}, both exploring builds from the CI process. \citet{ID126:RahmanRoy2017} focused on the influence of the status and frequency of these builds on code review, while \citet{ID768:ZampettiEtAl2019} targeted build failures and how these builds are discussed during code review. The key findings of these investigations indicate that passed builds and frequently built projects are more likely to encourage the reviewers' participation. In general, the status of a build influences the acceptance of a review request, but there are exceptions, and failed builds are sometimes accepted. Moreover, they found many review discussions on difficulties in configuring the CI pipeline.

\citet{ID538:PanichellaEtAl2015} examined what has been changed in the code during a review to understand how static analysis tools could have helped. By analyzing the changes during code reviews, \citet{ID538:PanichellaEtAl2015} found that 6\%--22\% of the warnings detected by static analysis approaches are removed during code checking. The analysis also indicates a trend of developers to focus on particular kinds of problems during a review, such as imports and regular expressions. Therefore, both studies of \citet{ID126:RahmanRoy2017} and \citet{ID538:PanichellaEtAl2015} suggest that other practices focusing on code quality might be helpful for code review, promoting participation and reducing the burden during the code checking.

Exploring a more specific topic, \citet{ID785:PaixaoEtAl2019} mined review data to investigate the intent and awareness of developers of the architectural impact of their changes, and also how architectural changes evolve during code review. As result, only 31\% of the examined reviews with a noticeable impact on the architecture have a conversation related to such impact, which suggests a lack of awareness when such type of modification is performed. However, \citet{ID785:PaixaoEtAl2019} also found that developers tend to be more often aware of the architecture when the change is related to it. In reviews in which the architecture is discussed, there is a trend to have larger improvements in cohesion and coupling.

The fifth study in this group is concerned with code ownership heuristics and whether code review data might complement them. \citet{ID267:ThongtanunamEtAl2016} analyzed how the code authoring and reviewing contributions differ to investigate whether the review activity should be used in the code ownership heuristics. By examining data of two OSS projects, the researchers found that 67\%--86\% of the developers are review-only contributors, being 18\%--50\% of them documented as core team members. In addition, there is evidence of an increasing relationship between the proportion of reviewers who have both low traditional and review ownership values with the likelihood of having post-release defects. This suggests that the reviewing activity can be used to refine the code ownership heuristics.

\begin{framed}
 \noindent\textbf{Finding 15}: Few warnings raised by static analysis tools are removed during code review. Contexts relying on continuous integration, with frequently built projects and automated builds with the passed status, are related to increased reviewers' participation. 
\end{framed}

 \subsection{RQ-1 - What foundational body of knowledge has been built based on studies of MCR?}
 The foundational body of knowledge of MCR consists mainly of evidence about the practice in real settings. The adoption and execution of MCR are influenced by multiple factors, being frequent findings related to developers' experience. The experience with other teams, projects, products, tools, and familiarity with the source code is reflected in the decisions that shape the review process, internal outcomes, and how reviewers behave when checking the code. MCR is adopted using different tools and rules in both industry and OSS. Despite the MCR variants, there are convergent refinements that remain over the years, such as the small number of reviewers and review comments per review request. As expectations, the review practitioners desire a positive impact on the code quality, knowledge sharing, and learning. The \textsc{foundational studies} provide evidence that MCR increases the code quality and promotes discussions about quality aspects, security, and architecture. Moreover, MCR influences the peers' perception, the developers' role in open-source projects, and code ownership. The foundational body of knowledge of MCR is commonly based on data analysis from historical information of open-source repositories, but there are also studies in large companies.


\section{\textsc{Proposals}}\label{sec:Proposal}

The previously discussed studies provide an understanding of how MCR works, deriving knowledge that is helpful to develop novel approaches to support MCR. This section introduces 53 \textsc{proposals} of a technique, tool, or theory that aim at improving the MCR practice. These are grouped into three categories, shown in Table~\ref{tab:proposalCodes}. The first two categories are related to the two phases of MCR described in Section~\ref{sec:Background}, namely \emph{Review Planning and Setup} and \emph{Code Review}. The third, \emph{Process Management and Support}, includes approaches that focus on aiding the MCR process by providing guidance, data analysis or tool support. We next discuss the identified \textsc{proposals}.

\begin{table*}[t]
\caption{Classification of MCR \textsc{Proposals}.}
\label{tab:proposalCodes}
\centering
\begin{tabular}{l l r}
\toprule
\textbf{Category} & \textbf{Proposal Goal} & \textbf{\#Proposals} \\ 
\midrule
\multicolumn{2}{l}{\textbf{Review Planning and Setup}} & \textbf{18} \\
\quad Patch Documentation & Assist the preparation of the review request & 1 \\
\quad Reviewer Recommender &  Recommend or automate the selection of reviewers & 14 \\
\quad Review Prioritization & Support the prioritization of review requests & 3 \\
\hline
\multicolumn{2}{l}{\textbf{Code Review}} & \textbf{19} \\
\quad Code Checking	& Support the activity of checking the code performed by reviewers & 15 \\
\quad Feedback Provision & Support the activity of providing feedback to authors & 2 \\
\quad Review Decision & Support deciding whether there is a need for further review & 2 \\
\hline
\multicolumn{2}{l}{\textbf{Process Management and Support}} & \textbf{16} \\
\quad Methodology and Guidelines	& Provide a taxonomy or guidelines to support MCR & 4 \\
\quad Review Retrospective & Assess or predict high-level MCR outcomes & 6 \\
\quad Tool Support &  Presentation of tools to support the MCR lifecycle & 7 \\
\bottomrule
\end{tabular}
\end{table*}

\subsection{Review Planning and Setup}

We identified three kinds of support associated with the \emph{Review Planning and Setup} phase, i.e.\ before reviewers review the code. The types of provided support are: (i)~\emph{patch documentation}: helping authors to complement the code change with information that is helpful to the review; (ii)~\emph{reviewer recommender}: aiding the selection of suitable reviewers; and (iii)~\emph{review prioritization}: helping reviewers to select code reviews to be performed earlier.

\subsubsection{Patch Documentation}\label{sec:patchDocumentation}

A single approach~\citep{ID224:HaoEtAl2013} is dedicated to support patch documentation. It consists of a tool---named Multimedia Commenting Tool (MCT)---implemented as an Eclipse plug-in. MCT allows programmers to include code narration and embedded multimedia resources, as well as support the replay of these comments. Thus, reviewers can reproduce them, which might help the understanding of code changes. MCT works with multimedia comments that contain audio or a video clip, a file recording mouse movements, a copy of the source file when the comment is created, and optionally extra files due to code changes.

\subsubsection{Reviewer Recommender} \label{sec:reviewerRecommenders}

Reviewer recommenders consist of tools and underlying techniques suggesting a list of the best candidates to review a review request. The motivation of most techniques is to reduce of the time taken for a review acceptance. There are also techniques whose goal is to support newcomers to reach out experienced developers~\citep{ID401:ZanjaniEtAl2016,ID155:RahmanEtAl2016}, to find the best candidates when the review request involves multiple files and large changes~\citep{ID600:OuniEtAl2016,ID493:XiaEtAl2017}, or to quickly perform recommendations at scale~\citep{ID706:AsthanaEtAl2019}.
These recommender techniques use as input the review request and complementary data from repositories of software development. The rationale behind these algorithms is to assign a score for reviewer candidates based on attributes of the historical databases, presenting as recommendation those with higher ratings. Some techniques also consider a time prioritization factor to give more weight to current than past reviews~\citep{ID051:YuEtAl2016, ID079:JiangEtAl2017, ID138:Balachandran2013, ID210:ThongtanunamEtAl2014, ID401:ZanjaniEtAl2016, ID493:XiaEtAl2017, ID638:FejzerEtAl2018,ID803:JiangEtAl2019}. A single study proposes a recommender with load balancing to mitigate the effect of unbalanced recommendations~\citep{ID706:AsthanaEtAl2019}.

The identified technique that was first published is called Review Bot~\citep{ID138:Balachandran2013}, which recommends as reviewers those who worked on the same lines of the review request. \citet{ID210:ThongtanunamEtAl2014, ID464:ThongtanunamEtAl2015} then suggested the use of \emph{file patch similarity}, which takes into account previous changes with similar paths and who reviewed them. Their second work presented RevFinder~\citep{ID464:ThongtanunamEtAl2015} was adopted as a baseline technique in the evaluation of most of the approaches proposed posteriorly. 
Other techniques explore a wide range of additional features to improve code recommenders, they are: (i) the \emph{review description} and its similarity with past descriptions~\citep{ID616:XiaEtAl2015}; (ii) the \emph{review feedback} written by potential reviewers~\citep{ID051:YuEtAl2016, ID401:ZanjaniEtAl2016}; (iii) common \emph{interests} among reviewers~\citep{ID600:OuniEtAl2016, ID493:XiaEtAl2017,ID772:LiaoEtAl2019}; (iv) cross-project experience in specialized \emph{technologies}~\citep{ID155:RahmanEtAl2016}; and (v) \emph{topic models} using review request information and reviewers' influence~\citep{ID772:LiaoEtAl2019}. Part of this information complemented those previously explored. \citet{ID616:XiaEtAl2015} extended RevFinder, while other approaches~\citep{ID600:OuniEtAl2016, ID155:RahmanEtAl2016} include the expertise with files to be reviewed. \citet{ID706:AsthanaEtAl2019} then prioritized who either committed changes or reviewed the files in the review request of a project. Instead of proposing a new technique, other studies investigated the effectiveness of different features to identify the best set of reviewers. Jiang \citet{ID079:JiangEtAl2017} compared the performance of the use of activeness, text similarity, file similarity, and social relation, concluding that activeness outperforms the others. 
Finally, \citet{ID638:FejzerEtAl2018} suggested building profiles of individual programmers for recommending a reviewer. This profiles can be updated as new reviews are made, so that it is not necessary to process past information for a recommendation. 

\subsubsection{Review Prioritization}

Given that reviewers might be invited to many code reviews, there is a need for prioritization. Thus, the approaches in this group provide support by suggesting which requests should be reviewed first. \citet{ID637:Aman2013} gives as recommendation a list of source files to be reviewed, using a 0-1 programming model-based method. The approach estimates the bug-proneness and the cost required for review each file. The goal is to find an optimal selection of files to be reviewed that does not exceed a certain cost while maximizes the sum of bug-proneness in this set. \citet{ID639:FanEtAl2018}, instead, suggested the estimation of the \emph{chances for a change to be accepted or rejected} in the code review process. First, a model building phase takes as input a set of labeled changes, extracts 34 features, and train a prediction model. In a prediction phase, this model can be applied to estimate if a change is going to be accepted. The idea is to give higher priority to high-quality changes. Lastly, \citet{ID753:WenEtAl2018} suggested directing the review effort considering the \emph{impact on the project deliverables}. The proposed tool, BLIMP Tracer, first extracts the building dependency graph of the system. Then, given the review request, the tool recursively traverses the graph to identify the impacted product deliverables. The output is an impact analysis report and the idea, in this case, is to give higher priority to changes impacting critical project deliverables or deliverables that cover a broad set of products.

\begin{framed}
 \noindent\textbf{Finding 16:} Approaches to support authors of code changes focus mainly on the \emph{Review Planning and Setup} phase of MCR (Figure~\ref{fig:overviewMCR}). Reviewer recommender is the most common type of support for the MCR process (14 of 53 \textsc{proposals}), to help the author select reviewers. Most of these techniques use a review (or development) history to build a recommendation. For the \emph{Preparation} task, there is tool support that allows authors to include multimedia comments on the source code, which might help reviewer understanding.
\end{framed}

\subsection{Code Review}

A set of approaches focuses on helping reviewers in the manual activity of analyzing a code change and providing comments. These approaches are split into three categories---code checking, feedback provision and review decision---which are discussed next. 

\subsubsection{Code Checking}

Support for code checking has been focused on providing information to ease the manual work of reviewers, mainly the understanding of a change. To provide this support, the main strategies presented in the studies are (i) provide visualizations of code changes~\citep{ID146:BarnettEtAl2015,ID641:FreireEtAl2018,ID234:TaoKim2015,ID400:GeEtAl2017,ID175:ZhangEtAl2015,ID276:DuleyEtAl2010,ID711:HuangEtAl2018,ID716:WangEtAl2019,ID817:GuoEtAl2019,ID736:HuangEtAl2018}, (ii) present properties associated with them~\citep{ID601:MenariniEtAl2017,ID405:TymchukEtAl2015,ID598:MishraSureka2014}, and (iii) support the analysis of change impact~\citep{ID614:WangEtAl2017,ID761:HanamEtAl2019}.

Some \textsc{proposals} assume that a change can be decomposed to ease understanding. ClusterChanges~\citep{ID146:BarnettEtAl2015}, JClusterChanges~\citep{ID641:FreireEtAl2018}, CoRA~\citep{ID716:WangEtAl2019}, and ChgCutter~\citep{ID817:GuoEtAl2019} are techniques for decomposing changes based on static analysis. ClusterChanges and JClusterChanges, instantiated for C\# and Java language, respectively, analyze composite changes to uncover definitions and their uses, clustering diff-regions into trivial and non-trivial. Similarly, CoRA analyzes dependency relationships and similarity in tangled changes, further generating a description of each partition based on its importance, templates, and program analysis technique. ChgCutter, in turn, partitions a user-selected change subset based on dependencies and builds a compilable intermediate program version,  in which a subset of regression tests can be run. In addition to this techniques, \citet{ID234:TaoKim2015} built a heuristic-based approach that identifies and groups two changed lines as related if (i) both are formatting-only changes; or (ii) they are semantically related for having static dependencies, or (iii) they are logically related for having similar change patterns.

Differently, there are approaches that are specific to particular kinds of change or language type. ReviewFactor~\citep{ID400:GeEtAl2017} focuses on separating \emph{refactoring} from non-refac\-toring changes, based on the code before and after the change and the log files of the refactoring tool usage. They take into account the automatic and manual refactorings, providing as output a visualization of the non-refactoring part and then the refactoring part. CRITICS~\citep{ID175:ZhangEtAl2015} targets the inspection of systematic changes, allowing authors to customize a change template to summarize similar changes. This is used to detect potential mistakes. ISC~\citep{ID711:HuangEtAl2018} focuses on identifying a salient class in a review request, which is a modified class that causes the modification of the remaning classes in the request. The problem is modeled as a binary classification using a large number of discriminative features. ClDiff~\citep{ID736:HuangEtAl2018} then works to generate concise linked code differences exploring the abstract syntax tree (AST). As output, ClDiff provides a visualization of code differences grouped by the links found and also a description of each group. The other more specific approach consists of a differencing algorithm, named Vdiff~\citep{ID276:DuleyEtAl2010}, which focuses on a hardware description language, Verilog. As typical diff tools assume sequential execution semantics, they are not suitable to hardware design descriptions, and thus need alternatives to identify changes. 

The approaches discussed above focus on \emph{syntactic} aspects of code changes. Other approaches that aim to support code checking target on the software behavior, quality, change impact analysis, and effort. Getty~\citep{ID601:MenariniEtAl2017} aims to aid code review with inter-version semantic differential analysis, presenting summaries of both code differences and \emph{behavioral differences}, using invariants extracted from the execution of test cases. Visual Design Inspection (ViDI)~\citep{ID405:TymchukEtAl2015} gives a city-based code visualization to help reviewers inspect the impact of changes on the overall \emph{quality} of the software. This visualization together with critics (broken design rules) are used by reviewers to indicate changes to be made. 
SemCIA (for JavaScript) and MultiViewer are assistance tools for change impact analysis. The former implements novel semantic relations and shows relationships between structural changes and changes to program behavior. MultiViewer~\citep{ID614:WangEtAl2017}, in turn,  includes the formal definition of three metrics: effort, risk, and impact. It presents this information in a Spider Chart and a Coupling Chart to support reviewers to identify coupling relations among related files in the changes. Concerning effort, \citet{ID598:MishraSureka2014} provide an estimation model for code review. Six variables to measure the size and complexity of the modified files are used to help reviewers predict the work needed to review a code change.

\subsubsection{Feedback Provision}

In addition to analyze code changes, reviewers must be able to provide feedback to authors, which can be, e.g., request for changes, ask questions and clarification, or votes of acceptance or rejection. Two approaches have the goal of supporting this feedback provision. Rich Code Annotation (RCA)~\citep{ID235:PriestPlimmer2006} is a digital ink tool integrated with the development environment to support annotations in reviews, so that reviewers can provide feedback using multimedia resources. The other approach consists of a prediction model---RevHelper~\citep{ID207:RahmanEtAl2017}---to indicate the usefulness of review comments in review submissions. This model was built on top of studies~\citep{ID131:BosuEtAl2015,ID207:RahmanEtAl2017} (discussed in Section~\ref{sec:Foundational}) on the usefulness of reviews comments (one of them published in the same paper in which RevHelper was proposed).

\subsubsection{Review Decision}

After going through a round of review, a code change can be accepted, rejected or may need rework, possibly requiring further reviews. To help reviewers reach a decision, there are two approaches that provide reviewers with complementary information that serve as indicators of whether a code change should be accepted. Both approaches have the goal to predict fault proneness. \citet{ID439:HarelKantorowitz2005} estimate the number of faults remaining in code. The method is an adaptation of an estimator (used in the formal software inspection process) to a scenario of iterative code review, where there are multiple review iterations. \citet{ID258:SoltanifarEtAl2016}, in turn, proposed a prediction model similar to typical bug predictors, which builds a model to predict fault proneness based on a set of features. The difference of their approach is that they consider features associated with the review that has been done to predict whether a patch remains defective.

\hfill \break \break \break

\begin{framed}
 \noindent\textbf{Finding 17:} Most MCR approaches focus on supporting reviewers, being 22 of 53 \textsc{proposals} (41.5\%) related to the reviewers' work, i.e., \emph{review prioritization}, \emph{code checking}, \emph{feedback provision}, and \emph{review decision}. From these 22 \textsc{proposals}, 15 target the \emph{code checking} task, aiming to help understand code changes, which is a challenge for practitioners, as identified in \textsc{foundational studies}. In 66.7\% of cases (10 of 15 studies), this checking support consists of providing alternative visualizations of code changes. 
\end{framed}

\subsection{Process Support}

The two groups of approaches previously described target specific tasks of MCR. We now detail the last group, which focuses on the MCR process as a whole and is split into three sub-groups.

\subsubsection{Methodology and Guidelines}

Existing works classified as methodology and guidelines are those that, based on collected data and previous studies, propose a taxonomy or guidelines to improve MCR. Taxonomies were proposed by \citet{ID414:BaumEtAl2016} and \citet{ID642:LiEtAl2017}. The former presented a faceted classification scheme for industrial MCR processes, including variations on how the process is embedded, reviewers aspects, code checking, feedback, and overarching facets. The latter consists of a taxonomy of topics in the review feedback, with four main categories (code correctness, pull-request decision-making, project management, and social interaction). 

Two works proposed guidelines. The first~\citep{ID655:BaumSchneider2016} makes recommendations associated with tool support, suggesting improvements in code review tools to increase review efficiency and effectiveness. The second study~\citep{ID413:BaumElAl2017} is a middle-range theory to indicate an optimal order to read the code, deriving six principles that define a proper order of changes and how they should be presented for reviewers.

\subsubsection{Review Retrospective}

MCR repositories can be explored to improve its process in particular projects. Six works went to this direction. \citet{ID291:UwanoEtAl2006} presented an integrated environment to capture eye movements during the code review, the Crescent tool. The approach uses an eye mark tracker, associating this information with a line of the code. This data is then available for further analysis, which allows developers to examine individual performance objectively. 

The other proposals explored the comments written by reviewers, presenting approaches to categorize them by means of machine learning algorithms. The studies aim to inform comment usefulness, confusion content, sentiment analysis, and identification of review topics. \citet{ID486:PangsakulyanontEtAl2014} proposed a semantic similarity classification of comment usefulness, in which the approach computed its semantic similarity with the review request description and observing if it satisfies threshold values. \citet{ID131:BosuEtAl2015} then proposed a classification based on eight comment attributes, such as the number of participants and comments in a thread, and the number of iterations in the review. These attributes were defined based on the findings of a preliminary exploratory study, in which developers reported their perception of usefulness. As the usefulness classifiers, the approach proposed by \citet{ID465:EbertEtAl2017} aimed to understand the content of review feedback. However, in this case, the researchers focused on the presence of confusion, grounded on the assumption that confusion negatively affects the effectiveness of code review. Based on an existing theoretical framework for categorizing expressions of confusion, eight different classifiers were trained with manually labeled the data, allowing an automatic confusion identification. Performing sentiment analysis in review comments, SentiCR~\citep{ID409:AhmedEtAl2017} is a supervised sentiment analysis tool designed explicitly for code review. It uses a sentiment oracle built empirically. Finally, \citet{ID642:LiEtAl2017} developed a two-stage hybrid classification of review topics. Grounded on its \textsc{foundational study}, the approach classifies the contents of review comments, allowing them to identify what reviewers are talking about. The intent is to organize the process and optimize the review tasks.

\subsubsection{Tool Support}

MCR is supported by widely used tools, such as Gerrit and CodeFlow. However, there are different tools that have been developed with similar purpose~\citep{ID132:MullerEtAl2012, ID378:KalyanEtAl2016, ID425:NagoyaEtAl2005, ID434:LanubileMallardo2002, ID572:ZhangEtAl2011, ID608:PerryEtAl2002, ID195:SripadaEtAl2016}. Table~\ref{tab:toolSupport} summarizes the approaches identified in our SLR (ordered by their publication date), their ultimate goal, and strategies adopted to achieve it. We list proposals to support the code review itself, as well as to support the development of code review tools. The latter includes a single work~\citep{ID195:SripadaEtAl2016} that consists of an extensible framework to the gamification of review systems, aiming to increase developers' interest. 

\begin{table*}
\caption{Summary of the approaches that provide tool support for MCR.}
\label{tab:toolSupport}
\centering
\small
\begin{tabular}{llll}
\toprule
\textbf{Approach} & \textbf{Year} & \textbf{Main Goal} & \textbf{Tool Style} \\ 
\midrule
\multicolumn{4}{l}{\textbf{Support Code Review}} \\
\quad IBIS~\citep{ID434:LanubileMallardo2002} & 2002 & Support to distributed code inspections & Web-based code inspection tool \\
\quad HyperCode~\citep{ID608:PerryEtAl2002} & 2002 & Support to distributed code inspections & Web-based code inspection tool \\
\quad Nagoya et al.~\citep{ID425:NagoyaEtAl2005} & 2005 & Support to the function-path review method & Desktop-based tool \\
\quad Java Sniper~\citep{ID572:ZhangEtAl2011} & 2011 & Promotion of collaborative code review & Web-based code review tool \\
\quad SmellTagger~\citep{ID132:MullerEtAl2012} & 2012 & Improvement in desirability and collaboration & Tool for tablet \\
\quad Fistbump~\citep{ID378:KalyanEtAl2016} & 2016 & Overcoming of limitations of existing tools & Web-based tool integrated with GitHub \\ 
\hline
\multicolumn{4}{l}{\textbf{Support the Development of Review Tools}} \\
\quad Sripada et al.~\citep{ID195:SripadaEtAl2016} & 2016 & Increase in the  developers' interest & Extensible framework for gamification \\
\bottomrule
\end{tabular}
\end{table*}

Older supporting tools refer to code inspections, proposing process changes to enable a more lightweight practice. This is the case of IBIS~\citep{ID434:LanubileMallardo2002} and HyperCode~\citep{ID608:PerryEtAl2002}, which can be used within a flexible and asynchronous review process. In contrast, recent tools have been focusing on more specific goals, aiming to address specific concerns of MCR, such as SmellTagger for tablets~\citep{ID132:MullerEtAl2012}.

\subsection{RQ2 - What approaches have been developed to support MCR?}

Multiple approaches to support MCR have been proposed. Most of them consist of reviewer recommenders and techniques to provide visualizations of code changes. Reviewer recommenders focus on finding suitable candidates to review a code change, commonly to reduce the time taken for a review acceptance. These recommenders are usually based on historical data of code review and consider the file path similarity or common interests among reviewers in the past to rank and suggest the reviewers' candidates to a new review request. Approaches that provide visualizations of code changes focus on helping the reviewers during code checking, highlighting change aspects to ease the understanding and the finding defects. The main strategy to provide visualizations is to decompose a code change based on static analysis, presenting the differences between the old and new versions of the decomposed code.


\section{\textsc{Evaluations}} \label{sec:Evaluation}

We now focus on how the approaches to support MCR tasks have been evaluated either individually or compared to a baseline. We first introduce the types of evaluation that appeared either in papers that include a \textsc{proposal} (discussed in the previous section) or in those that has as main contribution an evaluation of existing approaches. Then we detail design aspects of performed evaluations, followed by a discussion of their key findings.

\subsection{Evaluations Types}

\begin{figure}[t]
\includegraphics[trim=20 0 0 0,clip,width=\linewidth]{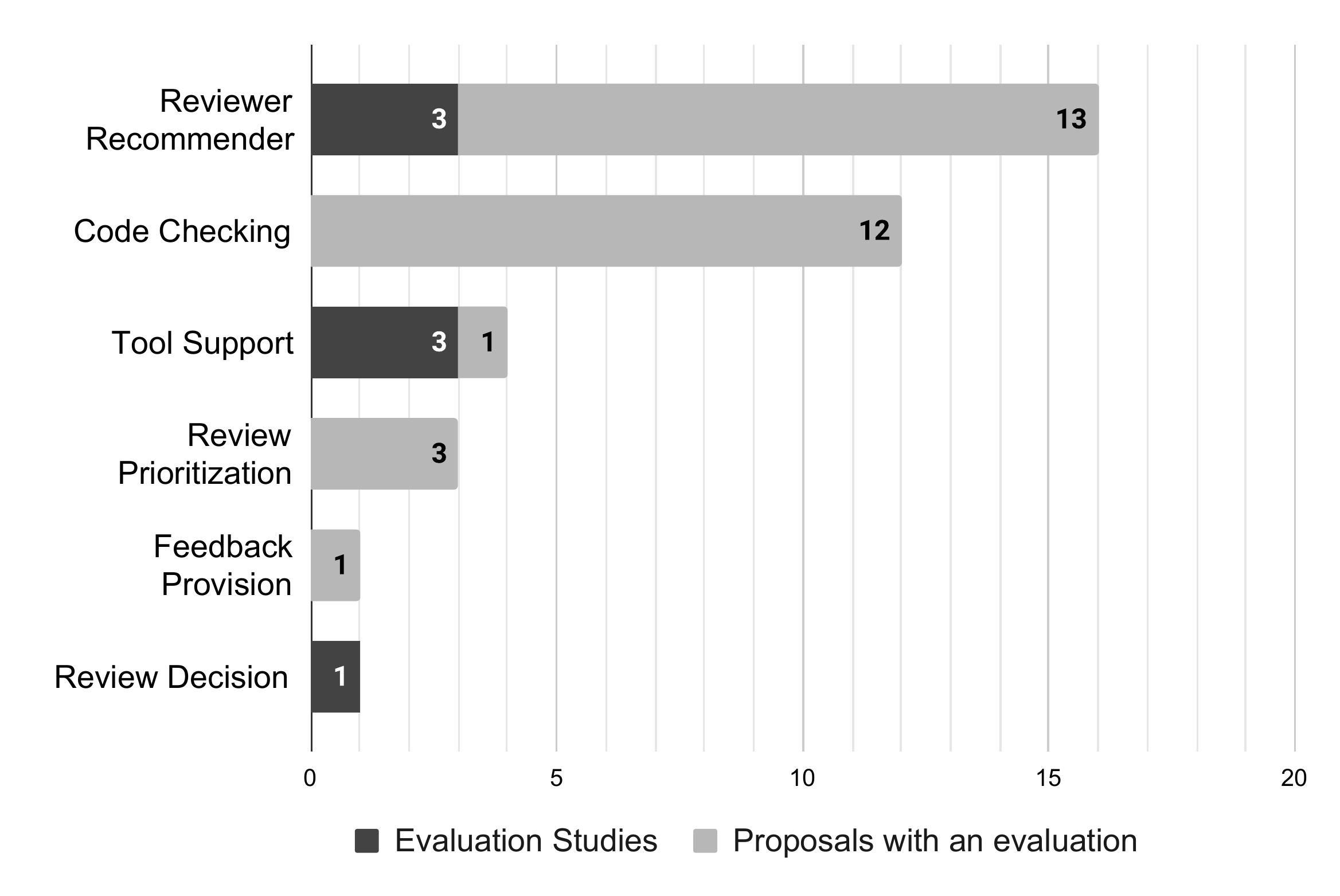}
\caption{Number of \textsc{evaluation} papers per type of MCR approach.} 
\label{fig:evaluationPresence}
\end{figure}

Considering the identified \textsc{proposals}, we investigated whether they were evaluated in the paper that they were proposed. From the 53 \textsc{proposals}, 30 (56.6\%) include an evaluation. We further identified seven papers whose main contribution is one or more evaluations. As result, there are 37 papers labeled as containing an \textsc{evaluation}. Figure~\ref{fig:evaluationPresence} shows the number of publications containing a \textsc{proposal} with an accompanying evaluation and evaluation studies. Most of these studies aim to assess a reviewer recommender technique, which is the most frequent type of approach. Typically, a newer approach is compared to the current state-of-the-art approach to demonstrate that it provides improvements at least in one aspect. Any form of evaluation of approach to support MCR was considered. We did not, however, consider as an evaluation a description of a scenario to illustrate the use or the benefits of a \textsc{proposal}, that is, when there is solely an example of its use made by its own authors; or the report of informally received feedback. This only provides anecdotal evidence of the effectiveness of the proposed approach.

Studies comparing MCR outcomes with other techniques, e.g.~pair programming, are not included in this section because they are considered \textsc{foundational studies} (Section~\ref{sec:Foundational}). In addition, we also do not include studies that consist of an analysis of different sets of features (i.e.\ feature selection) or learning algorithm to predict MCR outcomes, e.g.~\citep{ID079:JiangEtAl2017}. These are classified as \textsc{proposals}, as their contribution is an identified set of features/algorithm. The process of assessing the accuracy of different alternatives is considered part of the development of the approach.

\begin{figure}[t]
\includegraphics[trim=20 0 0 0,clip,width=\linewidth]{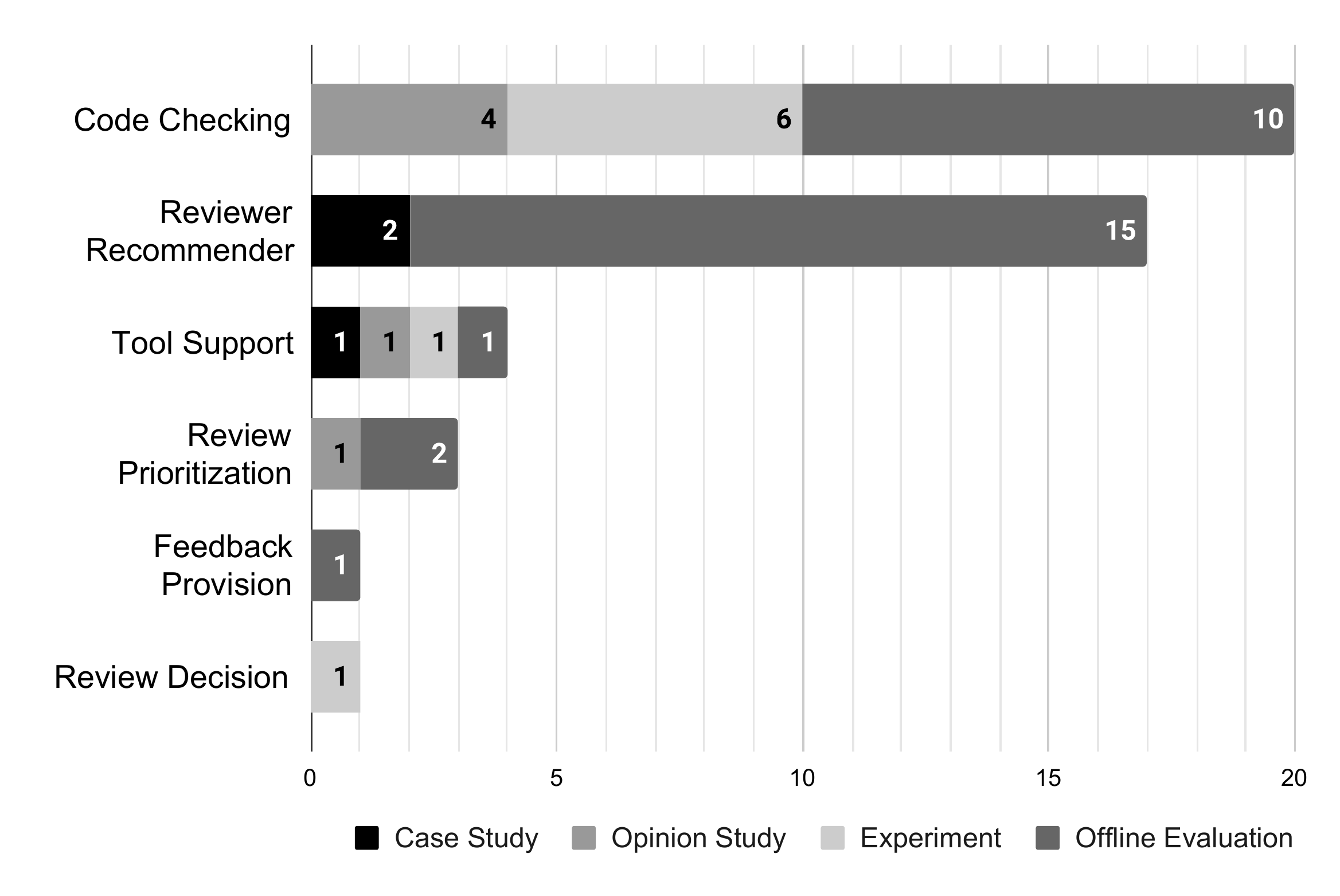}
\caption{Number of studies categorized by evaluation type and MCR approach type.} 
\label{fig:evaluationTypes}
\end{figure}

We classified the adopted research methods of \textsc{evaluations} into four main groups, as described in Table~\ref{tab:evaluationTypes}. The sum of the studies per type is higher than the number of papers because there are papers that include more than one type of evaluation study. The most common is \emph{offline evaluation}, being adopted in more than half (64.64\%) of the cases. This type of evaluation refers to studies in which researchers execute a technique or tool using existing data from MCR repositories as ground truth, and also as input of the approach, when it is the case. These offline evaluations do not involve human subjects. In Figure~\ref{fig:evaluationTypes}, we further detail the adopted evaluation type by the categories of MCR approaches. As can be seen, almost all reviewer recommenders have been evaluated offline.

\begin{table*}[t]
\caption{Classification of MCR \textsc{Evaluations}.}
\label{tab:evaluationTypes}
\centering
\begin{tabularx}{\textwidth}{l X r}
\toprule
\textbf{Category} & \textbf{Evaluation Description} & \textbf{\# Studies} \\ 
\midrule
Offline Evaluation & Evaluation of an MCR approach (with or without baseline) using historical data from software projects to validate the output of the approach. & 29 \\
Experiment & Empirical study in controlled settings to observe the effects of an MCR approach. & 7 \\
Opinion Study & Subjective (qualitative or quantitative) evaluation of an MCR approach by subjects, after introducing the approach and allowing participants to experiment it. & 6 \\
Case Study & Observation and collection of data from the instantiation of the approach in real settings, possibly using mixed research methods. & 3\\
\bottomrule
\end{tabularx}
\end{table*} 

Possibly due to the time and effort required to conduct user studies, evaluations involving human subjects, either professionals or students, were the choice in only a few studies. \emph{Experiments} are the second most frequent evaluation type in our review, reported in seven papers (15.91\%). These studies are characterized by a controlled environment, which involves participants and measured variables to analyze the effect of an intervention in the code review process. In contrast to experiments, \emph{opinion studies} have none or limited control, being the participants invited to a hands-on trial to try the proposed tool or technique. From this interaction with the proposed approach, the researchers collect the user perception using interviews or questionnaires. Therefore, the evaluation is based on collected subjective data. Studies of this type are present in six (13.64\%) papers. From these, four~\citep{ID146:BarnettEtAl2015, ID175:ZhangEtAl2015,ID817:GuoEtAl2019,ID716:WangEtAl2019} were used to complement the results of another study detailed in the same publication. One study~\citep{ID132:MullerEtAl2012} consists of a preliminary evaluation in which the subjects interact with a prototype of the proposed approach. Then, in another opinion study~\citep{ID753:WenEtAl2018}, the participants were invited to a semi-structured interview, in which they used the proposed approach and provided feedback. Lastly, \emph{case studies} collect and analyze data from a particular non-controlled environment. Three studies fall into this category, two of them based on OSS projects, and one from industry. In one of them, Peng et al.~\cite{ID281:PengEtAl2018} evaluated both the usage and perception of the developers on a reviewer recommender in GitHub, collecting quantitative data and interviewing developers. In the other case study, Mizuno and Liang~\cite{ID661:MizunoLiang2015} assessed the evolution of Gerrit using multiple sources, such as an interview with a developer of the tool and a comparison between Gerrit and Rietveld regarding features and code review logs. Finally, one case study~\citep{ID706:AsthanaEtAl2019} involved the deployment of the proposed approach in five repositories, following a collection of metrics to evaluate the usage and also a user study to analyze improvements. Later, this last case study was also complemented with an offline evaluation.

We next investigate in-depth the two most common types of evaluation, namely offline evaluations and experiments, detailing their designs and reached conclusions.

\subsection{Offline Evaluations}

Focusing on offline evaluations, we first discuss their study design in terms of (i) object of the study, i.e.\ evaluated approach; (ii) number and type of target projects; and (iii) metrics collected for the evaluation. We then examine the  conclusions reached by these studies.

\subsubsection{Study Design}

\paragraph{Study Object}

From the 29 offline evaluations, most (51.72\%) focuses on reviewer recommenders. Approaches to support code checking are the second most common (34.48\%)~\citep{ID146:BarnettEtAl2015,ID234:TaoKim2015,ID276:DuleyEtAl2010,ID400:GeEtAl2017,ID641:FreireEtAl2018,ID711:HuangEtAl2018,ID716:WangEtAl2019,ID761:HanamEtAl2019,ID817:GuoEtAl2019,ID736:HuangEtAl2018}. From the remaining works, two target review prioritization support~\citep{ID637:Aman2013,ID639:FanEtAl2018}, and the last two studies evaluate feedback provision assistance~\citep{ID207:RahmanEtAl2017} and a variation in the MCR process~\citep{ID257:BaumEtAl2016}.

Reviewer recommenders are usually evaluated using historical datasets, which are used to train and test generated models. The key steps typically followed in this kind of evaluation are: (i)~retrieving review requests with closed status; (ii) cleaning and sorting the collected data in chronological order; (iii) using part of the data to build a model using a learning technique; and (iv) using this model to make predictions in the test set and evaluating the results with a particular metric, such as precision, recall, and mean squared error (MSE).  

Approaches to support code checking, usually by means of visualizations, are also evaluated using historical datasets. The approaches are applied to existing changesets, and through a manual inspection the correctness of the output is verified, e.g., whether generated change clusters are acceptable. The manual inspection of the output followed two strategies, namely with and without a ground truth. Five approaches~\citep{ID234:TaoKim2015,ID276:DuleyEtAl2010,ID711:HuangEtAl2018,ID716:WangEtAl2019,ID817:GuoEtAl2019} to manipulate changeset visualizations were evaluated using a baseline created by human evaluators. \citet{ID234:TaoKim2015} included the first author and two external students as evaluators, while \citet{ID276:DuleyEtAl2010} and \citet{ID817:GuoEtAl2019} did not mention an external member in this step. \citet{ID711:HuangEtAl2018} and \citet{ID716:WangEtAl2019}, in turn, involved only external students in analyzing the existing data to manually establish the ground truth. In the three other studies, the output visualization was manually scrutiny by the researchers without a ground truth, using existing data of review requests as support. For example, \citet{ID146:BarnettEtAl2015} examined commit messages to verify the proposed partition of a changeset. Lastly, in two evaluations~\citep{ID761:HanamEtAl2019,ID736:HuangEtAl2018}, the output of the proposed approach is compared to another tool. Four of these papers related to support code checking~\citep{ID146:BarnettEtAl2015,ID234:TaoKim2015,ID817:GuoEtAl2019,ID736:HuangEtAl2018} include studies with human subjects to complement the offline evaluation.

The remaining four offline evaluations followed different procedures. \citet{ID637:Aman2013} produced review plans by the so-called ``conventional method'' and the proposed method, analyzing both recommendations by the number of buggy files included in the suggested list. In contrast, \citet{ID639:FanEtAl2018} and \citet{ID207:RahmanEtAl2017} proposed prediction models. Consequently, their evaluations consist of comparisons with other baselines. Moreover, \citet{ID207:RahmanEtAl2017} manually built a ground truth to evaluate the feedback assistance approach, while \citet{ID639:FanEtAl2018} used the stored data of a change request as a baseline for analysis. Finally, \citet{ID257:BaumEtAl2016} used a simulation model to analyze the differences between pre-commit review and post-commit review in terms of quality, efficiency and cycle time.

\paragraph{Target Projects}

Offline evaluations, in our context, use data from existing software projects. We detail in Table~\ref{tab:offlineSampleNumber} descriptive statistics of the number of the projects analyzed in each study and whether these projects are open source or proprietary. Two studies~\citep{ID155:RahmanEtAl2016, ID401:ZanjaniEtAl2016} are taken into account in both table rows as they used data from both types of projects. One study~\cite{ID257:BaumEtAl2016} is not considered in this table because the paper only mentions the use of data from the industry, without detailing information from which projects data was collected.

\begin{table*}[t]
\caption{Descriptive Statistics of the Target of Offline Evaluations and Experiments.}
\label{tab:offlineSampleNumber}
\centering
\begin{tabular}{l l r r r r r r}
\toprule
\textbf{Evaluation Type} & \textbf{Study Target} & \textbf{\#Studies} & \textbf{Mean}  & \textbf{SD}  & \textbf{Median}  & \textbf{Min}  & \textbf{Max}\\ 
\midrule
\multirow{2}{*}{Offline Evaluation} & Open-Source Projects & 24 & 13.29 & 28.22 & 4 & 1 & 120\\ 
& Proprietary Projects &  6  & 3.83   &  3.43 &  3 & 1 & 10\\\hline
Experiment & Subjects &  8  & 34.25   &   60.21 &  13 & 8 & 183\\
\bottomrule
\end{tabular}
\end{table*}

The majority of the studies collected data only from OSS repositories. The most frequently used projects are Android and OpenStack with seven occurrences each, followed by Qt and LibreOffice, which were adopted in six and four evaluations, respectively. Three offline evaluations~\citep{ID051:YuEtAl2016, ID281:PengEtAl2018,ID711:HuangEtAl2018} using open-source data did not inform which repositories they mined. Only four studies~\citep{ID138:Balachandran2013, ID146:BarnettEtAl2015, ID207:RahmanEtAl2017,ID706:AsthanaEtAl2019} used data solely from industry. 

With respect to the number of projects, most of the studies had as target only a few projects---the median is 4 projects that are open source and 3 projects that are proprietary. As an outliers, \citet{ID051:YuEtAl2016} and \citet{ID711:HuangEtAl2018} used data from 84 and 120 projects, respectively. Note that some of the projects are large scale and, therefore, contain a large amount of code review data, with many review requests and contributors, which can justify the low number of projects in some of the studies.

\paragraph{Metrics}

Approaches that rely on learning techniques use the metrics that are typically used in this context. Usually, the reported metrics are accuracy, precision, and recall. Particular studies also consider F-measure~\citep{ID051:YuEtAl2016, ID401:ZanjaniEtAl2016, ID638:FejzerEtAl2018,ID706:AsthanaEtAl2019,ID772:LiaoEtAl2019} or effectiveness ratio~\citep{ID639:FanEtAl2018}, for example. As reviewer recommendation can also be seen as a raking problem, another frequent evaluation metric is Mean Reciprocal Rank (MRR)~\citep{ID155:RahmanEtAl2016, ID464:ThongtanunamEtAl2015, ID600:OuniEtAl2016, ID616:XiaEtAl2015, ID638:FejzerEtAl2018,ID716:WangEtAl2019,ID803:JiangEtAl2019}. In particular, \citet{ID803:JiangEtAl2019} evaluated their proposed recommender considering the model construction time and prediction time.

Approaches that have a specific purpose elaborate custom metrics: (i) \citet{ID637:Aman2013} analyzed the number of buggy files in the generated output; (ii) \citet{ID400:GeEtAl2017} considered the refactoring ratio detected by the approach to assess its impact; (iii) \citet{ID638:FejzerEtAl2018} examined the memory footprint; (iv) \citet{ID257:BaumEtAl2016} used specific heuristics for quality, efficiency and cycle time;  and (v) \citet{ID761:HanamEtAl2019} analyzed the number of correct dependencies created by the proposed approach in comparison to others. Lastly, \citet{ID716:WangEtAl2019} evaluated the clustering result using Rand Index and \citet{ID736:HuangEtAl2018} examined the generation of the proposed visualization using accuracy, conciseness, and time performance.

\subsubsection{Findings}

Offline evaluations are based on quantitative data analysis. Therefore, reached conclusions indicate how an MCR approach performs and whether it outperforms an existing approach. We discuss key findings of the these evaluations by study object as follows.

\paragraph{Reviewer Recommenders}

Most of the papers on reviewer recommenders (15 out of 16) report results of an offline comparison between a proposed approach and selected baselines. Consequently, the main result is an evidence that indicates that the proposed approach is better than an existing one (selected baseline) according to a selected metric. Typically, the studies consider the top-k recommendations (where k is the number of recommended reviewers). We detail in Table~\ref{tab:findingsRecommendersComparisons} the reported results when a new approach is compared to another algorithm of reviewer recommendation, showing which approach outperformed which baseline according to which metric. 

\begin{table*}[t]
\caption{Results of comparisons of code reviewer recommenders. Table cells indicate when an approach (rows) outperformed a baseline listed in its corresponding column, with respect to a particular metric. The metrics are accuracy (ACC), mean reciprocal rank (MRR), and precision and recall (P\&R).}
\label{tab:findingsRecommendersComparisons}
\centering
\renewcommand{\tabcolsep}{2mm}
\small
\begin{tabular}{l|c|c|c|c|c|c|c|c}
\toprule
\multirow{2}{*}{\textbf{Approach}}          & \multicolumn{7}{c}{\textbf{Outperformed Reviewer Recommenders}} \\ \cline{2-9} 
                                            & Review Bot & FPS & RevFinder & TIE & cHRev & IR+CN & Activeness & WRC \\ \hline
FPS~\citep{ID210:ThongtanunamEtAl2014}       & ACC &&&&&&& \\
RevFinder~\citep{ID464:ThongtanunamEtAl2015} & ACC &&&&&&& \\
TIE~\citep{ID616:XiaEtAl2015}                &&& ACC &&&&& \\
Correct~\citep{ID155:RahmanEtAl2016}         &&& ACC, P\&R, MRR &&&&& \\
cHRev~\citep{ID401:ZanjaniEtAl2016}          &&& P\&R &&&&& \\
RevRec~\citep{ID600:OuniEtAl2016}            & P\&R && P\&R && P\&R &&& \\  
WRC~\citep{ID298:HannebauerEtAl2016}         && ACC &&&&&& \\
PR-CF~\citep{ID493:XiaEtAl2017}              && P\&R && P\&R && P\&R & P\&R& \\
\citet{ID638:FejzerEtAl2018}   & P\&R && P\&R &&&&& \\
TRFPre~\citep{ID803:JiangEtAl2019}&&&&ACC&ACC&&&ACC\\
WhoDo~\citep{ID706:AsthanaEtAl2019} &&&&&P\&R&&& \\
\bottomrule
\end{tabular}
\end{table*}

Other studies---presented in Table~\ref{tab:findingsRecommenders}---compared a reviewer recommender with alternative baselines, e.g.\ a simple heuristic or a standard learning technique. For instance, \citet{ID138:Balachandran2013} developed RevHistRECO, which is a heuristic inspired by the observed manual process of the reviewer assignment, adopted as a baseline for his proposed Review Bot. \citet{ID051:YuEtAl2016}, in turn, extended and implemented as baseline recommenders based on existing techniques, such as information retrieval (IR). Some of the baseline approaches in this table, e.g.\ xFinder~\citep{xFinder:KagdiEtAl2008} and CoreDevRec~\citep{CoreDevRec:Jiang2015}, are not in our SLR, because their purpose is not to recommend code reviewers and are not in the context of MCR.

\begin{table*}[t]
\caption{Results of comparisons between a reviewer recomender and a baseline technique. Each row indicates that an approach in the first column outperformed baselines in the third column, with respect to the metrics listed in the second column. The metrics are accuracy (ACC), F-measure (F1), mean reciprocal rank (MRR), and precision and recall (P\&R).}
\label{tab:findingsRecommenders}
\centering
\begin{tabularx}{\textwidth}{llX}
\toprule
\textbf{Approach} & \textbf{Measurement}                  & \textbf{Outperformed Baselines}                                                                                  \\ \midrule
Review Bot~\citep{ID138:Balachandran2013}  & ACC                 & RevHistRECO \\
IR+CN~\citep{ID051:YuEtAl2016} & P\&R and F1       & SVM-based, IR-based, FL-based, IR-based+CN-based, and FL-based+CN-based  \\
cHRev~\citep{ID401:ZanjaniEtAl2016}        & P\&R, F1, and MRR & xFinder and RevCom \\
WRC Algorithm~\citep{ID298:HannebauerEtAl2016} & ACC             & Line 10 Rule, Number of Changes, Expertise Recommender, Code Ownership, Expertise Cloud, and Degree-of-Authorship \\
TRFPre~\citep{ID803:JiangEtAl2019} &ACC& CoreDevRec and ACRec \\
\bottomrule
\end{tabularx}
\end{table*}

\paragraph{Code Checking}

While reviewer recommenders are usually compared to similar approaches, eight out of ten offline studies~\citep{ID146:BarnettEtAl2015, ID234:TaoKim2015, ID276:DuleyEtAl2010, ID400:GeEtAl2017, ID641:FreireEtAl2018,ID711:HuangEtAl2018,ID716:WangEtAl2019,ID817:GuoEtAl2019} that evaluate work to support code checking use a manually built ground truth. \citet{ID761:HanamEtAl2019}, however, used additional tools to build a ground truth. In both cases, it is used to identify false positives and false negatives given as output of the approaches to support code checking.  The identified false positives are then analyzed by the researchers, who examine their cause so that the proposed approach can be improved. For instance, \citet{ID146:BarnettEtAl2015} and \citet{ID641:FreireEtAl2018} highlighted which type of code change was not considered by their approach to distinguish trivial and non-trivial changes. The other offline evaluation~\citep{ID736:HuangEtAl2018} of a code checking approach then mixed the manual analysis of output and the comparison with an existing tool. As key findings, \citet{ID736:HuangEtAl2018} found that the proposed approach has higher accuracy and better conciseness as well as required less time in the study. 

\begin{framed}
 \noindent\textbf{Finding 18:} Evaluations of approaches to support MCR mainly focus on the validation of the output using data analysis. This strategy---named \emph{offline evaluation}---is usually used when assessing reviewer recommenders and code checking support. Most offline evaluations use data from OSS projects. Precision and recall are the most frequent metrics in this type of evaluation. Considering reviewer recommenders, RevFinder has been the most used as a baseline.
\end{framed}

\begin{framed}
 \noindent\textbf{Finding 19:} Most conclusions of the \textsc{evaluation} studies are related to the feasibility and effectiveness of an approach to support MCR based on data analysis. The results of the evaluation of a newly-developed proposal outperforms baselines, achieving better accuracy, precision, and/or recall.
\end{framed}

\subsection{Experiments}

A smaller amount of MCR approaches (in comparison to offline evaluations) have been evaluated by means of experiments. They are performed in controlled environments and involve subjects, being them students, professionals, or both. We next discuss the following aspects of the design of these studies: (i) independent and dependent variables; and (ii) participants. As in the previous section, we also summarize their findings.

\subsubsection{Study Design}

\paragraph{Variables of the experiments}

The experimental studies to evaluate an MCR approach rely on various variables. Five of these studies adopted a within-subjects design, in which all participants performed review tasks using both the proposed approach and another selected for comparison. In the experiment of \citet{ID175:ZhangEtAl2015}, the analysis of systematic changes in the code was performed using the proposed CRITICS and also, as baseline, the diff and search features of Eclipse. Similarly, \citet{ID761:HanamEtAl2019} asked the study participants to perform a review task supported by the proposed change impact analysis tool SemCIA and also by other tools, SynCIA and UnixDiff. In two studies, the participants reviewed a code change with and without the support of the proposed approach output, i.e.\ with or without partitions~\citep{ID234:TaoKim2015} and with or without a hint of salient class~\citep{ID711:HuangEtAl2018}. Lastly, participants of the study of  \citet{ID736:HuangEtAl2018} conducted the review with the proposed tool ClDiff and with GumTree, the state-of-the-art tool to generate fine-grained code differences. In all these cases, the researchers measured the correctness of the output of the review task and the time spent. More specifically, in two of these studies~\citep{ID711:HuangEtAl2018,ID736:HuangEtAl2018} the experiment analyzed the degree of understanding the changes.

In contrast with these studies, two evaluations followed a between-subjects design. \citet{ID102:KhandelwalEtAl2017} evaluated the effect of gamification on code review, organizing the participants into five groups, each of them using either a gamified or a non-gamified review tool. The researchers then measured the subject's interest by the number of review comments, the usefulness of review comments, the number of identified bugs, the number of identified code smells, and the time spent. In ~\citet{ID601:MenariniEtAl2017}'s experiment, the intervention group used the proposed Getty approach, while the control group used GitHub resources. From these interactions, the code review process resulting from the used supporting approach has been assessed.

Finally, \citet{ID698:RunesonWohlin1998} performed an experiment to compare three alternative capture-recapture methods, which are used to estimate the number of bugs in a code after going through review. Participants had to review a target code and point out bugs. Based on this, the researchers analyzed the identified bugs and inspected the errors in the estimation of the evaluated methods.

\paragraph{Sample size}

The experiments performed to evaluate MCR approaches considered varying numbers of subjects to participate in the studies. In Table~\ref{tab:offlineSampleNumber}, where we show the number of projects used in offline evaluations, we also detail the descriptive statistics of the number of subjects in experiments. The study that involved the highest number of subjects (183) was conducted by \citet{ID102:KhandelwalEtAl2017}, while \citet{ID698:RunesonWohlin1998} experiment involved the smallest sample, with 8 subjects. Moreover, considering the background of subjects, three of the experiments~\citep{ID102:KhandelwalEtAl2017,ID175:ZhangEtAl2015,ID234:TaoKim2015} were conducted solely with students; the other three experiments~\citep{ID601:MenariniEtAl2017,ID698:RunesonWohlin1998,ID761:HanamEtAl2019} involved professionals in addition to students. In the experimented conducted by \citet{ID711:HuangEtAl2018}, the participants are characterized as people engaged in computer-related work with programming experience, but it is not clear whether they are professionals, students, or both.

\subsubsection{Findings}

Experiments performed in the context of MCR focus on evaluating particular aspects of the proposed approaches. Thus, the key findings target the value promoted by each approach. Two experiments~\citep{ID175:ZhangEtAl2015, ID234:TaoKim2015} provide evidence of a positive impact in the review process due to the proposed approach, such as by improving the correctness and time spent in the review activity. Similarly, three studies~\citep{ID711:HuangEtAl2018,ID761:HanamEtAl2019,ID736:HuangEtAl2018} give evidence of a specific positive impact of the proposed approach on the understanding of code changes. In contrast, \citet{ID102:KhandelwalEtAl2017} found no evidence that gamified tools promote a positive impact on the subjects' interest. Additionally, grounded on observations of the experiment execution, \citet{ID601:MenariniEtAl2017} indicated that semantically-assisted code review is feasible and effective.

In addition to the main collected data, three experiments also include a follow-up study with the participants to collect their perceptions about the proposed approach. By means of a survey~\citep{ID102:KhandelwalEtAl2017, ID175:ZhangEtAl2015} or an interview~\citep{ID601:MenariniEtAl2017}, the  subjective opinion of the participants suggests that the approaches might indeed help the code review process.

\subsection{RQ3 - How have MCR approaches been evaluated and what were the reached conclusions?}

Most of the approaches proposed to support MCR have been evaluated using offline studies, with the goal of validating the output of the technique or tool, commonly using historical data from code review as input. From 37 \textsc{evaluation} papers, 29 (78.38\%) present an offline study. In 12 of 29 cases, these validations consider the effectiveness by measuring precision and recall of the output compared to the stored information. In 10 of 29 offline studies, the validations consider the accuracy. The reached conclusions are usually related to a newly-developed approach that outperforms baselines by achieving better effectiveness in an offline evaluation. There are also few reports of \textsc{evaluations} with user studies (13 of 37 \textsc{evaluations}), and they are mainly focused on observing the effects and the opinion of the proposed approach.


\section{Discussion}
\label{sec:Discussion}

The literature on MCR reviewed in our SLR allowed us to identify and analyze the researched aspects of the practice, classify the existing approaches to support it, and understand how these supporting approaches have been evaluated. Thus, we presented in previous sections a structured body of knowledge of the MCR practice, which is helpful for both researchers and practitioners. In this section, we discuss further insights derived from our analysis.

\subsection{Historical Developments}

\begin{figure}[t]
\includegraphics[trim=20 0 0 0,clip,width=0.9\linewidth]{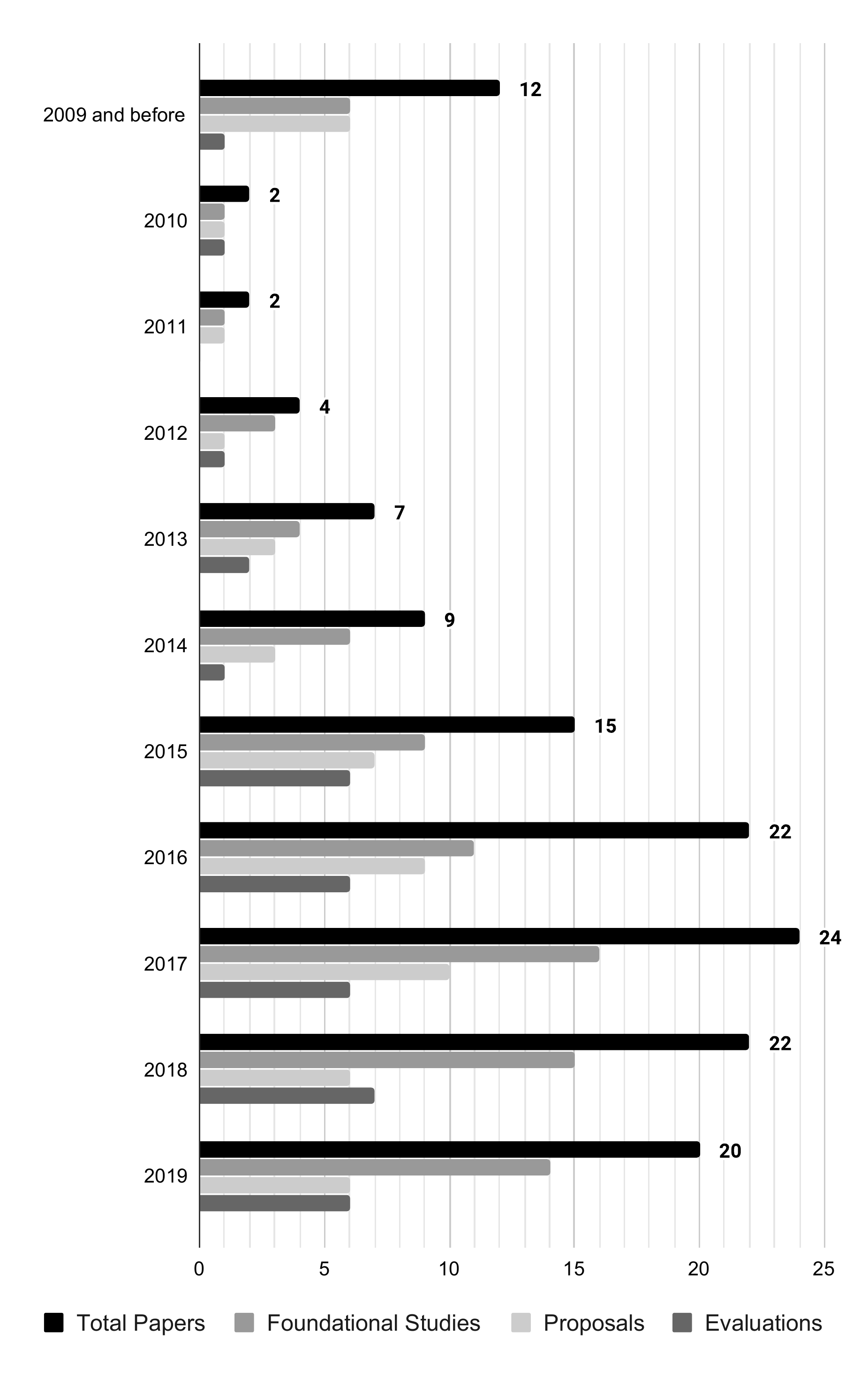}
\caption{Distribution of primary studies by type per year of publication.} 
\label{fig:studiesPerYear}
\end{figure}

Given the primary studies analyzed in this systematic review, we provide a historical perspective of the developments in the field of MCR in Figure~\ref{fig:studiesPerYear}. We summarize the total number of papers published per year, as well as papers of each type per year. We can observe an increase in research work on MCR in recent years. This trend indicates the importance and timeliness of the topic. Nevertheless, in the last two years (i.e., 2018 and 2019) we can observe a decrease in the total number of papers. This variation is mainly due to a decreased number of \textsc{proposals}. Nevertheless, this decline in the number of papers in the last couple of years does not provide enough evidence to claim any medium- and long-term trend. 

\subsection{Taxonomy of Research Work on MCR}

\begin{figure*}[t]
 \centering
 \includegraphics[width=0.85\linewidth]{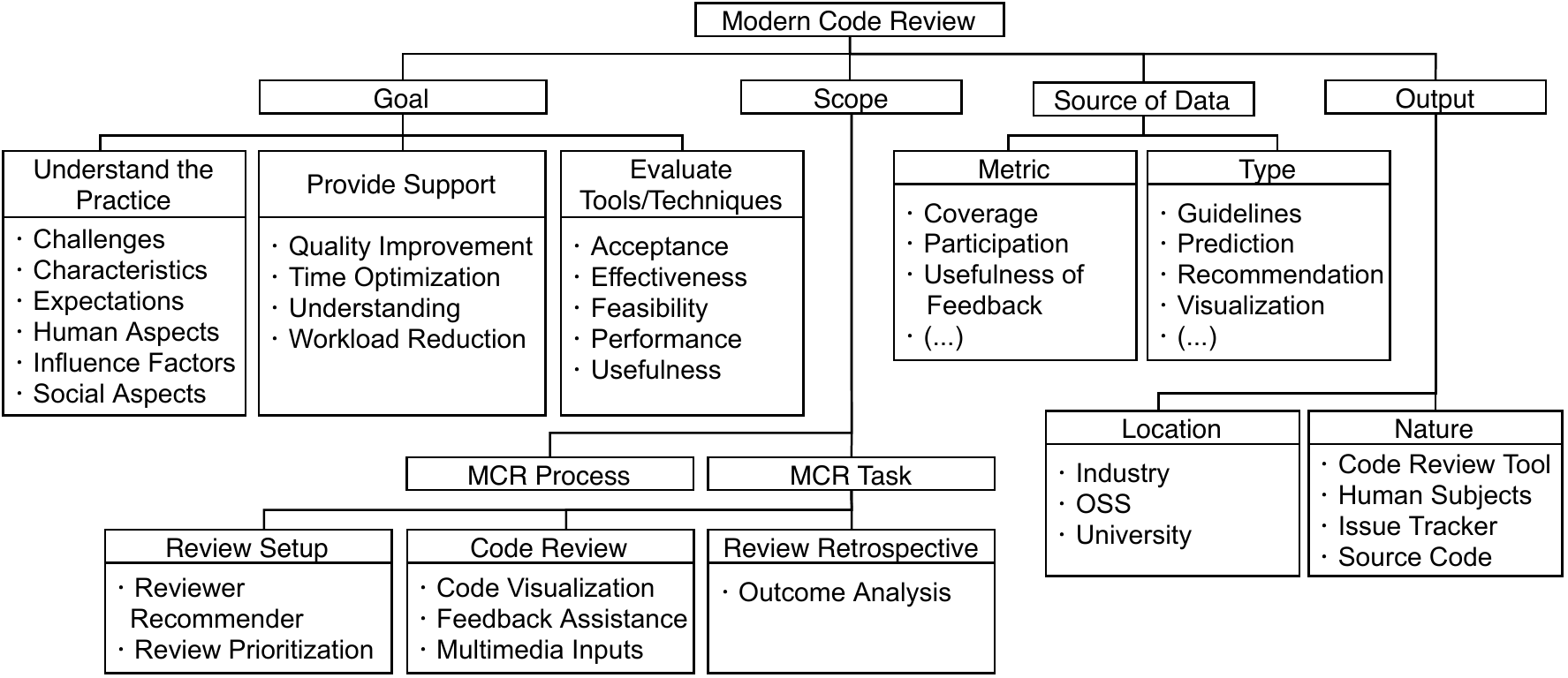}
 \caption{A taxonomy of MCR research based on primary studies of our SLR.}
 \label{fig:researchedAspects}
\end{figure*}

Our analysis of the primary studies of MCR selected for being investigated in this SLR leads us to identify key topics in this research area. We categorized the topics investigated in \textsc{foundational studies}, \textsc{proposals} of approaches to support the practice, and types of \textsc{evaluation} studies in the previous sections. We now summarize and propose a taxonomy of aspects of the research on MCR in Figure~\ref{fig:researchedAspects}, which complements the mentioned categories.

In our taxonomy, we first highlight the \emph{goals} of the research work on MCR, which can be related to: (i) a better understanding of one or more code review aspects, e.g.\ challenges faced by practitioners and factors influencing an outcome; (ii) novel approaches that provide support to practitioners; and (iii) evaluation techniques or tools, which aim to assess the effectiveness of MCR-supporting approaches. Works on MCR target a particular \emph{scope}, which can be the MCR process as a whole or a particular task. Studies or proposals generally rely on collected or existing data, which can be mined from repositories or obtained from human subjects. These data can be from a combination of locations, e.g.\ OSS or academia. This is captured by the \emph{source of data} facet. Finally, the last facet refers to the \emph{output} associated with the contribution of the work (which can be assessed metrics and/or a particular type of contribution).

\subsection{Actionable Implications and Future Directions}

The findings of our systematic literature review revealed that multiple aspects of MCR have been addressed in the last years. This section discusses the implications of our findings and issues that remain unaddressed.

\paragraph{\textbf{MCR process improvement}} Empirical studies of MCR demonstrate the feasibility of extracting information from the history of review activity stored in tools, such as Gerrit. For practitioners, these findings suggest how to improve internal outcomes, such as that the submission of small code changes increases reviewers' participation and reduces the review duration. Moreover, the findings related to the CodeFlow Analytics from Microsoft~\citep{ID158:BirdEtAl2015} demonstrate that developers can use review data to improve MCR. Although CodeFlow Analytics is an internal platform of Microsoft, there is an opportunity for researchers to study how developers can use this data to improve the MCR process in their own projects and what outcomes can support strategic decision making to improve this process.

\paragraph{\textbf{Code Improvement}} Differently from inspections, MCR has as one of its main benefits source code improvement. Nevertheless, changes based on comments are made in an individual basis. Consequently, the same comments might be made in different code reviews. Similar code improvements in reviewed requests have been identified~\citep{ID767:UedaEtAl2019} and common themes emerged from the manual analyses of review feedback~\citep{ID105:BacchelliBird2013, ID154:SpadiniEtAl2018}. Moreover, there is evidence that authors repeatedly introduce the same types of problems despite the reviewer feedback~\citep{ID758:UedaEtAl2018}. For practitioners, this implies more attention before submitting a request, disseminating and checking for issues that were earlier addressed, and reducing the review cost. For researchers, these findings suggest that natural language processing (NLP) techniques can be used to extract recurrent bad practices in projects based on comments to avoid them to occur again by means of knowledge dissemination.

\paragraph{\textbf{Exploration of Non-technical MCR Benefits}} Differently from code inspections, whose primary focus was bug detection, MCR brings various other benefits, such as knowledge transfer, collective code ownership, and learning. However, these non-technical benefits have been little explored in \textsc{foundational studies} and existing approaches do not focus on improving them.

\paragraph{\textbf{User studies on MCR}} Only few experiments involving human subjects have been conducted in the context of MCR. In our set of 139 primary studies, 19 reported an experiment (13.67\%). Including other methodological approaches involving participants, i.e., experiments, interviews, and surveys, 47 of 139 (33.81\%) papers present a study with practitioners or students. MCR is essentially a human-based activity, supported by tools. Consequently, further user studies must be done. Most of the evaluations of code reviewer recommenders rely only on accuracy metrics. However, as known in the recommender systems research area, other aspects, such as novelty and transparency, are key for the adoption of recommenders. Recently, \citet{MCR:UserStudies} observed this issue and conducted an \emph{in vivo} performance evaluation of a reviewer recommender. As a key finding, the researchers indicate the need for more user-centric approaches to designing and evaluating the recommenders. Therefore, user studies can help researchers to understand how approaches to support MCR are used and perceived and what are the barriers faced by practitioners.

\paragraph{\textbf{Studies in small and medium-sized companies}} From the 86 \textsc{foundational studies}, 55 studies selected data from 70 different open-source projects. The most frequent projects have been Qt, OpenStack, and Android with 23, 19, and 13 occurrences, respectively. In 17 of 55 studies, there is an overlap in which data from two or three of these projects were used in the same study, e.g., Qt and OpenStack are both data source projects in 13 studies. Considering research in the industry, our systematic review identified 17 papers involving data and participants from companies, a small number compared to studies involving open-source projects. From these studies in the industry, 6 are projects conducted at Microsoft. While the variety of open-source projects might suggest that the existing evidence of MCR spans multiple contexts, we observe as an implication the opportunity of future research to explore code review in other industrial contexts, such as small and medium-sized companies, to understand whether the existing knowledge might be generalized to other scenarios.

\subsection{Research Limitations} \label{sec:limitations}

The goal of this study is to identify the state of the art on MCR, providing a structured overview of the research done in this field. We thus performed a \emph{systematic} review in order to minimize the research bias, mitigating the influence of the researchers' expectations during the selection and analysis of a large number of primary studies.

To mitigate the limitations of our SLR in the review planning phase, we selected widely used digital libraries as sources and specified keywords as search string that cover the studied theme, assuming that the selected strategy would retrieve the largest number of relevant studies. Three of our selected databases---ACM Digital Library, IEEE Xplore, SpringerLink---were also used in other secondary studies on MCR~\citep{MappingMCR:BadampudiEtAl2019,MappingMCR:FronzaEtAl2020,MappingMCR:NazirEtAl2020}. While these other studies also searched the Scopus database, we selected ScienceDirect as a source. However, both Scopus and ScienceDirect are provided by Elsevier.

Although this search is large, it is not complete and, therefore, our review does not cover all existing work on MCR. An example of a study that was not retrieved in our search is that of \citet{RigbyEtAl2008} because it uses the term peer review, which is not covered by our search string. In this case, however, the findings of this study were used for comparison purposes by ~\citet{ID116:RigbyBird2013}, which is a primary study in our review. As discussed when we introduced our search string, some authors used \emph{peer review} to refer to the MCR practice, but this term is also used in other contexts, such as peer review of scientific papers. Our search identified papers that used the term \emph{peer code review}, but not solely \emph{peer review}. The advantage of systematic reviews is that they can be further extended in future reviews. 

Moreover, to identify further studies, a snowballing approach could have been used. This approach has not been followed in the present work because a preliminary analysis of the obtained results indicated that the most relevant studies had been retrieved by searching our selected digital libraries. Nevertheless, we estimated whether and how the lack of snowballing could have influenced our review. We randomly selected 20 papers from the set of primary studies, creating our referencing list. We then conducted a backward snowballing~\citep{SnowballingGuidelines:Claes2014} using this reference list to identify additional primary studies. We followed the same selection process of our SLR in this backward snowballing, using the same selection criteria. We checked 619 new results from our reference list and identified 8 papers that should be included in our review but were not identified by our procedure. From these eight papers, we checked 240 other results, including 2 papers. Then, we checked 26 results from those 2 papers, but we did not find more entries. Our procedure to estimate whether and how the lack of snowballing could have influenced our review identified 10 papers that were not identified by our search. The papers identified in this procedure used specific terms to refer to MCR practice, such as ``PR review''~\citep{SWB527:YingEtAl2016} and ``patch review''~\citep{SWB694:MehrdadEtAl2009}. Our estimation illustrates the amount of work required to perform backward snowballing. As our focus was conducting a systematic literature review, not a systematic mapping study, we suggest snowballing as future work, in which both backward and forward snowballing can be conducted as a complement to our review. In addition, our SLR in the present form already includes several papers. Including the snowballing approach could lead to a lengthy paper, which might be difficult to read.

The selection and classification of primary studies might also be considered a threat to validity. To avoid bias, we followed systematic procedures for the search, selection, data extraction, information labeling, and data analysis. Moreover, these tasks were conducted by the first author of this paper, and then the outcomes were reviewed and discussed with the second author. A single researcher analyzing the primary studies in an SLR is a practice observed in other studies, e.g. in the work of \citet{SLR:AccessibilitySE:PaivaEtAl2021}. However, these are subjective tasks, and other studies can result in different categorizations and summarizations of the research on MCR. 


\section{Conclusion}
\label{sec:Conclusion}

Code review is a well-known practice of quality assurance in software development that evolved from a structured and rigid form (i.e.\ software inspection) to a flexible, tool-based, and asynchronous process, namely modern code review (MCR). As MCR gained increasing popularity in recent years, the practice has been largely investigated in academia. Therefore, to have a comprehensive view of what has been done in this field, we presented in this paper the results of a systematic literature review on MCR, which includes \numPapers{} primary studies.

We identified three main categories of studies, namely \textsc{foundational studies}, \textsc{proposals} of novel approaches to support the practice, and \textsc{evaluations} of proposed approaches. Each paper category was systematically analyzed observing aspects relevant for each type of study. Most of the investigated work consists of \textsc{foundational studies} that have been conducted to better understand the motivations for the adoption of MCR, its challenges and benefits, and analysis of which influence factors lead to which MCR outcomes. From the \textsc{proposals} of novel approaches to support MCR, the most common are those to help code checking and to recommend reviewers. \textsc{Evaluations} of MCR-supporting approaches have been done mostly offline and few studies involving human subjects have been conducted.

We performed a \emph{systematic} literature review in order to reduce the bias in the selection and analysis of a large number of primary studies on MCR. Although this search is large, it is not complete and, therefore, our review may have left out other existing studies on MCR. To mitigate the limitations of SLRs, we selected widely used digital libraries as sources, assuming that they would contain the largest number of relevant studies. Future SLRs on MCR may target other digital libraries or gray literature as well as cover future years given that the research on MCR has been active to a great extent in the recent years.

\section*{Acknowledgements}
Ingrid Nunes thanks for CNPq grants ref. 313357/2018-8 and ref. 428157/2018-1. This study was financed in part by the Coordena\c{c}\~{a}o de Aperfei\c{c}oamento de Pessoal de N\'{i}vel Superior - Brasil (CAPES) - Finance Code 001.

\bibliography{main}
\end{document}